\documentclass[final,3p,times,twocolumn]{elsarticle}

\usepackage{ecrc}
\usepackage{lineno}

\volume{00}

\firstpage{1}

\journalname{Journal of Systems and Software}

\runauth{}

\jid{procs}

\jnltitlelogo{Procedia Computer Science}

\usepackage{amssymb}
\usepackage[figuresright]{rotating}

\usepackage{floatrow}
\newfloatcommand{capbtabbox}{table}[][\FBwidth]

\usepackage{amsmath,algorithm,hyperref,booktabs,algpseudocode,algc,pdfpages,caption,subcaption,listings,xcolor,amssymb,multirow,verbatimbox,fancyvrb,filecontents, graphicx}

\usepackage{verbatim,tikz}
\usepackage[table]{colortbl}
\usetikzlibrary{shapes.geometric, arrows, fit, positioning, decorations.markings, calc, snakes}
\graphicspath{ {img} }
\definecolor{codegreen}{rgb}{0,0.6,0}
\definecolor{codegray}{rgb}{0.5,0.5,0.5}
\definecolor{codepurple}{rgb}{0.58,0,0.82}
\definecolor{backcolour}{rgb}{0.95,0.95,0.92}

\lstdefinestyle{mystyle}{
    commentstyle=\color{codegreen},
    keywordstyle=\color{magenta},
    numberstyle=\tiny\color{codegray},
    stringstyle=\color{codepurple},
    basicstyle=\ttfamily\footnotesize,
    breakatwhitespace=false,         
    breaklines=true,                 
    captionpos=b,                    
    keepspaces=true,                 
    numbersep=0pt,                  
    showspaces=false,                
    showstringspaces=false,
    showtabs=false,                  
    tabsize=2
}
\newsavebox{\verbsavebox}

\makeatletter
\AtBeginDocument{\DeclareCaptionSubType{lstlisting}
\renewcommand{\p@sublstlisting}{\thelstlisting}

}
\makeatother
\newcommand{\zzh}[1]{{\color{black} #1}}

\newcommand{\zzhb}[1]{{\color{black} #1}}
\newcommand{\bo}[1]{{\color{black}  #1}}

\lstdefinestyle{MyListStyle} {
    numbers=left,
    basicstyle=\linespread{0.94}\ttfamily,
    language=C++
}

\def\BibTeX{{\rm B\kern-.05em{\sc i\kern-.025em b}\kern-.08em
    T\kern-.1667em\lower.7ex\hbox{E}\kern-.125emX}}

\begin{document}
%\linenumbers
\begin{frontmatter}

\title{Precise Learning of Source Code Contextual Semantics via Hierarchical Dependence Structure and Graph Attention Networks}

\author[mymainaddress]{Zhehao Zhao}
\ead{zhaozhehao@pku.edu.cn}

\author[mysecondaryaddress]{Bo Yang\corref{mycorrespondingauthor}}
\ead{yangbo@bjfu.edu.cn}

\author[mymainaddress]{Ge Li}
\ead{lige@pku.edu.cn}

\author[mythirdaryaddress]{Huai Liu}
\ead{hliu@swin.edu.au}

\author[mymainaddress]{Zhi Jin\corref{mycorrespondingauthor}}
\cortext[mycorrespondingauthor]{Corresponding author}
\ead{zhijin@pku.edu.cn}

\address[mymainaddress]{Key Laboratory of High Confidence Software Technologies, Peking University, Beijing 100871, China}
\address[mysecondaryaddress]{School of Information Science and Technology, Beijing Forestry University, Beijing 100083, China}
\address[mythirdaryaddress]{Department of Computing Technologies, Swinburne University of Technology, Hawthorn VIC 3122, Australia}

%% Title, authors and addresses

%% use the tnoteref command within \title for footnotes;
%% use the tnotetext command for the associated footnote;
%% use the fnref command within \author or \address for footnotes;
%% use the fntext command for the associated footnote;
%% use the corref command within \author for corresponding author footnotes;
%% use the cortext command for the associated footnote;
%% use the ead command for the email address,
%% and the form \ead[url] for the home page:
%%
%% \title{Title\tnoteref{label1}}
%% \tnotetext[label1]{}
%% \author{Name\corref{cor1}\fnref{label2}}
%% \ead{email address}
%% \ead[url]{home page}
%% \fntext[label2]{}
%% \cortext[cor1]{}
%% \address{Address\fnref{label3}}
%% \fntext[label3]{}

\dochead{}
%% Use \dochead if there is an article header, e.g. \dochead{Short communication}

%% use optional labels to link authors explicitly to addresses:
%% \author[label1,label2]{<author name>}
%% \address[label1]{<address>}
%% \address[label2]{<address>}

\address{}

\begin{abstract}
\zzhb{
Deep learning is being used extensively in a variety of software engineering tasks, e.g., program classification and defect prediction.
Although the technique eliminates the required process of feature engineering, the construction of source code model significantly affects the performance on those tasks.
Most recent works was mainly focused on complementing AST-based source code models by introducing contextual dependencies extracted from CFG. However, all of them pay little attention to the representation of basic blocks, which are the basis of contextual dependencies.

In this paper, we integrated AST and CFG and proposed a novel source code model embedded with hierarchical dependencies. Based on that, we also designed a neural network that depends on the graph attention mechanism.
Specifically, we introduced the syntactic structural of the basic block, i.e., its corresponding AST, in source code model to provide sufficient information and fill the gap. We have evaluated this model on three practical software engineering tasks and compared it with other state-of-the-art methods. The results show that our model can significantly improve the performance. For example, compared to the best performing baseline, our model reduces the scale of parameters by 50\% and achieves 4\% improvement on accuracy on program classification task.
}
\end{abstract}

\begin{keyword}
\texttt{Graph Neural Network}\sep 
\texttt{Program Analysis}\sep 
\texttt{Deep Learning}\sep 
\texttt{Abstract Syntax Tree}\sep
\texttt{Control Flow Graph}
\end{keyword}

\end{frontmatter}

%%
%% Start line numbering here if you want
%%
% \linenumbers

%% main text
\section{Introduction}\label{sec:1}
Recently, deep learning has been increasingly applied into program analysis tasks, such as program classification~\cite{Wang2019LearningBP,url3, url4}, software defect prediction~\cite{url1, url2}, and code summarization~\cite{url5, url6, url8}. 
\zzhb{However, the performance on these tasks heavily depends on the choice of source code model, which can be divided into three types: abstract syntax tree- (AST-) based, control flow graph- (CFG-) based and the hybrid model of these two. 
Moreover, depending on the structure of AST adopted during analysis, AST-based source code model can be further divided to the whole AST~\cite{url10, url9, url12} or partial AST~\cite{url13, c2v, c2s}. The syntactic structure within AST can illustrate all the information of source code, especially the subtle changes on it. However, the contextual dependencies are implicit in AST and cannot be extracted and learnt effectively. 
In contrast, the CFG-based source code model~\cite{url14, PDGICSE18} is good at providing contextual dependencies, which can be learnt effectively by graph neural networks. Nevertheless, CFG is uneffective to represent the information of statements located in the basic blocks. 
Therefore, some researches proposed methodologies to embed the contextual dependencies from CFG into AST\cite{url11, Li2019ImprovingBD, Alon2018AGP}.
Such a design idea of the hybrid method still take AST as the core part of the source code model. It would add the contextual dependencies as additional edges~\cite{url11} to AST or as assistant features~\cite{Li2019ImprovingBD}.
However, the basic blocks, which are the basis of contextual dependencies, are paid little attention by the existing methodologies.
To mine the contextual dependencies effectively, we argue that the features of basic blocks should be prioritized.}
%Thus, considering to propose a more precise model, we need to extract and represent information within basic blocks before introducing contextual dependencies into the source code model.
Figure~\ref{exp:simi} shows our motivational example. These two code segments come from the PROMISE dataset used in our study. The defect in Figure~\ref{exp:simi}(a) is that returning a \textit{null} value on line 7 will cause a \textit{NullPointerException}, and the corresponding fix is to return a \textit{Field} type array of length 0 here. After analyzing this example, we have the following observations.

\begin{figure}[!h]
\centering
\small
\begin{lrbox}{\verbsavebox}
\begin{lstlisting}[
        style=MyListStyle,
        xrightmargin=0.5\linewidth,
        firstnumber=1,
        ]
 class Document {
 
     public final Field[] getFields(String name){
         List result = new ArrayList();
         ...
         if(result.size() == 0)
             return null;
         ...
     }
 }
\end{lstlisting}
\end{lrbox}
\subcaptionbox{The defect version of \textit{lucene-2.2}\label{fig:1:a}}{
    \resizebox{0.4\linewidth}{!}{
    \usebox{\verbsavebox}
    }
}\hfill
\begin{lrbox}{\verbsavebox}
\begin{lstlisting}[
        style=MyListStyle,
        xrightmargin=0.5\linewidth,
        firstnumber=1,
        ]
 class Document {
     private final static Field[] NO_FIELDS = new Field[0];
     public final Field[] getFields(String name) {
         List result = new ArrayList();
         ...
         if(result.size() == 0)
             return NO_FIELDS;
         ...
     }
 }
\end{lstlisting}
\end{lrbox}
\subcaptionbox{The fixed version of \textit{lucene-2.4}\label{fig:1:b}}{
    \resizebox{0.4\linewidth}{!}{
    \usebox{\verbsavebox}
    }
}
\caption{A Motivating Example from PROMISE dataset}\label{exp:simi}
\end{figure}

\zzhb{
\textbf{\textit{Observation 1: This defect depends on the actual execution path.}} As shown in Figure~\ref{fig:1:a}, the defect is triggered only if the condition on line 6 is met. However, if the caller of the \textit{getFields} function properly handles caught exceptions, this defect will not be triggered. Thus, a reasonable source code model should reflect the execution path. Furthermore, since a large number of invocations to \textit{getFileds} are outside from the \textit{Document} class, the source code model should not be limited to a certain granularity.

\textbf{\textit{Observation 2: These two source codes differ slightly but with total different semantics.}}
As shown in Figure~\ref{exp:simi}, the difference of these two source codes is a choice between returning an identifier \textit{NO\_FIELDS} or a \textit{null} in line 7. The code in Figure \ref{fig:1:b} does not cause the exception because that \textit{NO\_FIELDS} refers to an \textit{object} (see line 2 of Figure \ref{fig:1:b}). Thus, the difference of these two source code is actually the difference between \textit{object} and \textit{null}. Moreover, for the deep learning models with some textual features (e.g., Bag of Words), the learning of these two words (\textit{null} and \textit{NO\_FIELDS}) is uneven, since \textit{null} occurs more frequently than \textit{NO\_FIELDS}, which would raise the difficulty for models to learn the real difference.

\zzh{According to the \textbf{Observation 1}, CFG would intuitively become the first choice of source code model, since CFG can show the potential execution path and can be constructed on any granularity. But there still exists the issue about how to represent the basic blocks within the CFG. In the existing CFG-based works, basic blocks are mainly represented by either line numbers~\cite{Li2019ImprovingBD} or Bag of Words~\cite{Zhong, ginn}. However, according to the \textbf{Observation 2}, these methods only utilize the textual features, which significantly relies on the frequency of occurrence. Thus they cannot effectively capture the difference shown in Figure~\ref{exp:simi} to distinguish \textit{NO\_FIELDS} and \textit{null}. We argue that a proper source code model should introduce semantic differences (e.g., the difference between \textit{object} and \textit{null}) into the deep learning models more than the textual distinctions.
}

Motivated by these observations, we propose a novel source code model.
Specifically, to overcome the limitation mentioned in the \textbf{Observations 1}, we choose CFG with dataflow (ECFG), which can reflect the actual execution paths, as the backbone of the source code model. To address the \textbf{Observations 2}, we use the block-level AST, i.e., each AST subtrees correspond to each ECFG basic blocks. Take the source codes in Figure~\ref{exp:simi} to illustrate the benefit of such way: since \textit{NO\_FIELDS} represents an \textit{object} while \textit{null} is just a keyword, the syntax rule for them are not same, which brings different AST structures.
To sum up, the whole model can be divided into two levels. At the outer level, we use the inter-procedure ECFG to express the dependencies between the basic blocks. At the inner level, we choose AST to express the structure of each basic block.

Our source code model has three advantages. First, benefiting from the ECFG as the main body, the granularity of our source code model can be flexibly adjusted. Second, also benefiting from the ECFG, our source code model can show the potential execution path explicitly, thus the contextual dependencies can be captured effectively by a graph neural network. Third, benefiting from the substructure of AST, our source code model can have a more informative representation of basic blocks, hence the features within each basic blocks can be captured effectively by a tree-based neural network. 

Furthermore, we designed a \textit{Multi-Flow Graph Neural Network} (MFGNN) to extract features from our source code model. The calculation of MFGNN can be divided into three steps. At the first step, we obtain features named \textit{local features} through TBCNN~\cite{url10} from the collection of AST-substructures, which are corresponded to the basic blocks in ECFG. 
At the second step, we extract features named \textit{contextual features} from ECFG, whose basic blocks has been filled with \textit{local features}. Since the ECFG is a directed graph with multi-typed edges where we want to adopt attention mechanism, we did a slightly modification on the original Graph Attention Network (GAT). Specifically, the modified model supports directed graph and multi-typed edges, we name it as \textit{Attention-based Graph Network for Directed Graph} (AGN4D), and apply it in the second step. 
At the third step, we apply a fusion layer to coalesce these features into hybrid features, which can be used for subsequent tasks.
}

To be specific, this paper has the following three major contributions:

\begin{itemize}
\item We propose a source code model that combines AST and \zzhb{CFG with dataflow (ECFG)}. The source code model can reflect both contextual dependencies and syntactic structure, which allows neural networks to learn richer program features.
%The combination graph can preserve the structural information in the program code as well as the syntactic information of each statement, which can facilitate the learning of precise semantics of the program.

\item We design a learning model to obtain contextual semantics from the source code model, namely Multi-Flow Graph Neural Network (MFGNN).
%, which can help to extract both structural and syntactic information. It is worth mentioning that MFGNN integrates an attention mechanism based layer for handling the combination graph.
\zzhb{MFGNN integrates an attention-based graph learning layer evolves from GAT.}

\item MFGNN is implemented and evaluated on three typical tasks, namely the program classification,  software defect prediction and code clone detection. 
The results show that MFGNN can extract richer program features than the state-of-the-art methods, and hence greatly improve the performance of these tasks.
%The results show that MFGNN can extract more precise contextual semantics for programs than the state-of-the-art approaches, and hence greatly improve the performance of these tasks.
\end{itemize}

The remainder of this paper is organized as follows: Section 2 introduces the background of our work. Section 3 describes the new source code model and MFGNN. We report our experimental studies and results in Sections 4 and 5, respectively. The related work is discussed in Section 6. Finally, we conclude this paper in Section 7.

\section{Background}
In this section, we would introduce some basic concepts and terms that are used in this paper.
\subsection{Program representation}
To represent a piece of program, there are several ways: token sequences, AST, CFG~\cite{PDGICSE18}. Among all of them, AST and CFG are adopted most widely, thus we would introduce both of them in this section.

\subsubsection{Abstract Syntax Tree}
\zzhb{
Abstract Syntax Tree (AST) is a tree representation of the abstract syntactic structure of source code written in a programming language~\cite{url10}. Each node on the AST represents a nonterminal symbol in the syntax rules of the programming language. Being a near-source-level program graph structure, AST can represent the syntactic information of programs in a simple way, which makes AST widely used in a variety of software engineering tasks\cite{url10,url11,url12,url13,url1,c2v,c2s}.}

\subsubsection{Control Flow Graph} Control Flow Graph (CFG) is a directed graph in which each node (namely basic block) represents a set of sequentially executed instruction sequences, and the edges represent control flow paths. CFG is mostly used in static analysis and compiler applications, as it can accurately represent the flow inside a program. For example, through graph reachability analysis, CFG can help locate inaccessible code in programs, and find syntax structures such as loops. As a source code model for deep learning, CFG's edges are usually considered to represent the contextual dependencies, which have a significant impact on the performance of software engineering tasks\cite{Li2019ImprovingBD,FCD,url11}.

\subsection{Graph Neural Networks}\label{sec:gat}
Graph is a generic data structure to effectively abstract objects and their connections~\cite{zhou2018graph}. It has been widely used across multiple domains, such as social networks~\cite{hamilton2017inductive}, chemical interaction~\cite{fout2017protein} and knowledge modeling~\cite{hamaguchi2017knowledge}. 

\zzh{Graph Neural Networks (GNNs) are methods used to mine the information within a graph and obtain the embedding vector of the graph under a learning model. GNNs are mostly based on the message-passing mechanism, and consist of two functions: the Message function and the Aggregate function~\cite{zhou2018graph}. The Message function is used to transform the original vector of nodes to obtain the hidden vector; and the Aggregate function is used to aggregate the transformed vectors of a node's adjacency nodes and obtain an embedding vector of the node. 

The Message function is generally represented using a parameter $W \in \mathbb{R}^{F*F'}$. Let $X = \{x_1, x_2,...,x_n\}, x_i \in \mathbb{R}^F$ be the initial features of nodes, and $H = \{h_1, h_2,...,h_n\}, h_i \in \mathbb{R}^{F'}$ be the transformed features of nodes. Then, the Message function can be defined as:
$$h_i = Message(x_i) = Wx_i$$
, where $F$ represents the initial dimension of nodes' features, and $F'$ represents the transformed dimension of nodes' features. 

Different GNNs often vary in the Aggregate functions. For example, GCN~\cite{GCN} uses summation as the Aggregate function, which is defined as follows.
$$h'_i = Aggregate(h, \mathcal{N}_i) = \sum_i^{\mathcal{N}_i} h_i$$
, where $\mathcal{N}_i$ is the collection of adjacency nodes of $i$.

GAT~\cite{url18} uses the self-attention mechanism as the Aggregate function. GAT first calculates \textit{self-attention} weights for all edges in the graph, as defined below:
$$\alpha_{ij} = \frac{exp(LeakyReLU(a(Wh_i||Wh_j)))}{\Sigma_{k\in\mathcal{N}_i}exp(LeakyReLU(a(Wh_i||Wh_k)))}$$
, where $||$ is the concatenation operation and
$a : \mathbb{R^{F'}} \times \mathbb{R^{F'}} \to \mathbb{R}$ is the shared attention mechanism.

GAT then linearly combines the transformed features of the neighbouring nodes according to the attention weights, which is defined as:
$$h'_i = \sigma(\sum_j^{\mathcal{N}_i}\alpha_{ij}h_j)$$
where $\sigma$ is a nonlinearity function.
}

\definecolor{beaver}{rgb}{0.62, 0.51, 0.44}
\definecolor{fuchsia}{rgb}{0.96, 0.0, 0.63}
\definecolor{orange}{rgb}{1.0, 0.31, 0.0}
\definecolor{mahogany}{rgb}{0.75, 0.25, 0.0}
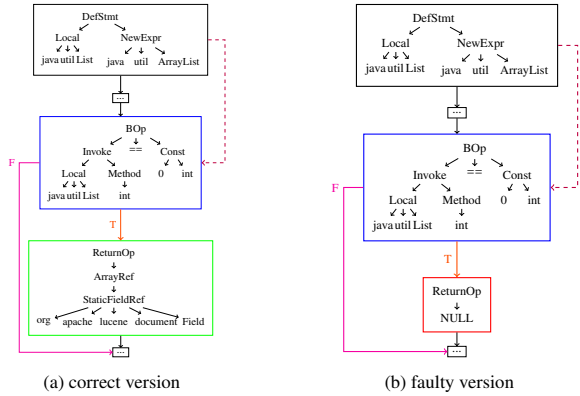
\begin{figure}[!h]
\centering
\subcaptionbox{correct version}{
    \resizebox{!}{0.6\linewidth}{
    \begin{tikzpicture}
        \tikzstyle{every node} = [font=\small]
        \node (ffast0) {DefStmt};
        \node[below left = 0.5em and -0.5em of ffast0] (ffast1) {Local};
        \node[below left = 0.5em and -1em of ffast1] (ffast2) {java};
        \node[below = 0.5em of ffast1] (ffast3) {util};
        \node[below right = 0.5em and -1em of ffast1] (ffast4) {List};
        \node[below right = 0.5em and -0.5em of ffast0] (ffast5) {NewExpr};
        \node[below left = 0.5em and -1em of ffast5] (ffast6) {java};
        \node[below = 0.5em of ffast5] (ffast7) {util};
        \node[below right = 0.5em and -1em of ffast5] (ffast8) {ArrayList};

        \draw[->] (ffast0) -- (ffast1);
        \draw[->] (ffast0) -- (ffast5);
        \draw[->] (ffast1) -- (ffast2);
        \draw[->] (ffast1) -- (ffast3);
        \draw[->] (ffast1) -- (ffast4);
        \draw[->] (ffast5) -- (ffast6);
        \draw[->] (ffast5) -- (ffast7);
        \draw[->] (ffast5) -- (ffast8);

        \node[draw, rectangle, thick, fit=(ffast0) (ffast1) (ffast2) (ffast3) (ffast4) (ffast5) (ffast6) (ffast7) (ffast8)] (ff0) {};

        \node[rectangle, draw, below = 1.5em of ff0, align=left] (ff1) {...};

        \node[below right = 1.5em and -0.6em of ff1] (ff2ast1) {BOp};
        \node[below left = 0.5em and 0.5em of ff2ast1] (ff2ast2) {Invoke};
        \node[below = 0.5em of ff2ast1] (ff2ast3) {==};
        \node[below right = 0.5em and 0.5em of ff2ast1] (ff2ast12) {Const};
        \node[below left = 0.5em and -1em of ff2ast2] (ff2ast4) {Local};
        \node[below right = 0.5em and -1em of ff2ast2] (ff2ast5) {Method};
        \node[below left = 0.5em and -1em of ff2ast12] (ff2ast6) {0};
        \node[below right = 0.5em and -1em of ff2ast12] (ff2ast11) {int};
        \node[below left = 0.5em and -1em of ff2ast4] (ff2ast7) {java};
        \node[below = 0.5em of ff2ast4] (ff2ast8) {util};
        \node[below right = 0.5em and -1em of ff2ast4] (ff2ast9) {List};
        \node[below = 0.5em of ff2ast5] (ff2ast10) {int};

        \draw[->] (ff2ast1) -- (ff2ast2);
        \draw[->] (ff2ast1) -- (ff2ast3);
        \draw[->] (ff2ast1) -- (ff2ast12);
        \draw[->] (ff2ast2) -- (ff2ast4);
        \draw[->] (ff2ast2) -- (ff2ast5);
        \draw[->] (ff2ast4) -- (ff2ast7);
        \draw[->] (ff2ast4) -- (ff2ast8);
        \draw[->] (ff2ast4) -- (ff2ast9);
        \draw[->] (ff2ast5) -- (ff2ast10);
        \draw[->] (ff2ast12) -- (ff2ast11);
        \draw[->] (ff2ast12) -- (ff2ast6);

        \node[draw, rectangle, blue, thick, fit=(ff2ast1) (ff2ast2) (ff2ast3) (ff2ast4) (ff2ast5) (ff2ast6) (ff2ast7) (ff2ast8) (ff2ast9) (ff2ast10) (ff2ast11) (ff2ast12)] (ff2) {};

        \node[below left = 3em and -8.4em of ff2] (ff3ast0) {ReturnOp};
        \node[below = 0.5em of ff3ast0] (ff3ast1) {ArrayRef};
        \node[below = 0.5em of ff3ast1] (ff3ast2) {StaticFieldRef};
        \node[below left = 0.5em and 2em of ff3ast2] (ff3ast3) {org};
        \node[below left = 0.5em and -1.5em of ff3ast2] (ff3ast4) {apache};
        \node[below = 0.5em of ff3ast2] (ff3ast5) {lucene};
        \node[below right = 0.5em and -1.5em of ff3ast2] (ff3ast6) {document};
        \node[below right = 0.5em and 2.5em of ff3ast2] (ff3ast7) {Field};

        \draw[->] (ff3ast0) -- (ff3ast1);
        \draw[->] (ff3ast1) -- (ff3ast2);
        \draw[->] (ff3ast2) -- (ff3ast3);
        \draw[->] (ff3ast2) -- (ff3ast4);
        \draw[->] (ff3ast2) -- (ff3ast5);
        \draw[->] (ff3ast2) -- (ff3ast6);
        \draw[->] (ff3ast2) -- (ff3ast7);

        \node[draw, green, rectangle, thick, fit=(ff3ast0) (ff3ast1) (ff3ast2) (ff3ast3) (ff3ast4) (ff3ast5) (ff3ast6) (ff3ast7)] (ff3) {};

        \node[rectangle, draw, below = 1em of ff3, align=left] (ff4) {...};

        \draw[->] (ff0) -- (ff1);
        \draw[->] (ff1) -- (ff2);
        \draw[->, thick, orange] (ff2) -- node[left] {T} (ff3);
        \draw[->] (ff3) -- (ff4);
        \draw[->, thick, fuchsia] (ff2.west) -- +(-0.6, 0) node[left]{F} |- (ff4);
        \draw[->, dashed, thick, purple] (ff0.east) -- +(0.5, 0) |- (ff2);

    \end{tikzpicture}
    }
}\hfill
\subcaptionbox{faulty version}{
    \resizebox{!}{0.6\linewidth}{
\begin{tikzpicture}
    \tikzstyle{every node} = [font=\small]
    \node (ffast0) {DefStmt};
    \node[below left = 0.5em and -0.5em of ffast0] (ffast1) {Local};
    \node[below left = 0.5em and -1em of ffast1] (ffast2) {java};
    \node[below = 0.5em of ffast1] (ffast3) {util};
    \node[below right = 0.5em and -1em of ffast1] (ffast4) {List};
    \node[below right = 0.5em and -0.5em of ffast0] (ffast5) {NewExpr};
    \node[below left = 0.5em and -1em of ffast5] (ffast6) {java};
    \node[below = 0.5em of ffast5] (ffast7) {util};
    \node[below right = 0.5em and -1em of ffast5] (ffast8) {ArrayList};

    \draw[->] (ffast0) -- (ffast1);
    \draw[->] (ffast0) -- (ffast5);
    \draw[->] (ffast1) -- (ffast2);
    \draw[->] (ffast1) -- (ffast3);
    \draw[->] (ffast1) -- (ffast4);
    \draw[->] (ffast5) -- (ffast6);
    \draw[->] (ffast5) -- (ffast7);
    \draw[->] (ffast5) -- (ffast8);

    \node[draw, rectangle, thick, fit=(ffast0) (ffast1) (ffast2) (ffast3) (ffast4) (ffast5) (ffast6) (ffast7) (ffast8)] (ff0) {};

    \node[rectangle, draw, below = 1.5em of ff0, align=left] (ff1) {...};

    \node[below right = 1.5em and -0.6em of ff1] (ff2ast1) {BOp};
    \node[below left = 0.5em and 0.5em of ff2ast1] (ff2ast2) {Invoke};
    \node[below = 0.5em of ff2ast1] (ff2ast3) {==};
    \node[below right = 0.5em and 0.5em of ff2ast1] (ff2ast12) {Const};
    \node[below left = 0.5em and -1em of ff2ast2] (ff2ast4) {Local};
    \node[below right = 0.5em and -1em of ff2ast2] (ff2ast5) {Method};
    \node[below left = 0.5em and -1em of ff2ast12] (ff2ast6) {0};
    \node[below right = 0.5em and -1em of ff2ast12] (ff2ast11) {int};
    \node[below left = 0.5em and -1em of ff2ast4] (ff2ast7) {java};
    \node[below = 0.5em of ff2ast4] (ff2ast8) {util};
    \node[below right = 0.5em and -1em of ff2ast4] (ff2ast9) {List};
    \node[below = 0.5em of ff2ast5] (ff2ast10) {int};

    \draw[->] (ff2ast1) -- (ff2ast2);
    \draw[->] (ff2ast1) -- (ff2ast3);
    \draw[->] (ff2ast1) -- (ff2ast12);
    \draw[->] (ff2ast2) -- (ff2ast4);
    \draw[->] (ff2ast2) -- (ff2ast5);
    \draw[->] (ff2ast4) -- (ff2ast7);
    \draw[->] (ff2ast4) -- (ff2ast8);
    \draw[->] (ff2ast4) -- (ff2ast9);
    \draw[->] (ff2ast5) -- (ff2ast10);
    \draw[->] (ff2ast12) -- (ff2ast11);
    \draw[->] (ff2ast12) -- (ff2ast6);

    \node[draw, rectangle, blue, thick, fit=(ff2ast1) (ff2ast2) (ff2ast3) (ff2ast4) (ff2ast5) (ff2ast6) (ff2ast7) (ff2ast8) (ff2ast9) (ff2ast10) (ff2ast11) (ff2ast12)] (ff2) {};

    \node[below = 3em of ff2] (ff3ast0) {ReturnOp};
    \node[below = 0.5em of ff3ast0] (ff3ast1) {NULL};

    \node[draw, red, rectangle, thick, fit=(ff3ast0) (ff3ast1)] (ff3) {};

    \node[rectangle, draw, below = 1em of ff3, align=left] (ff4) {...};

    \draw[->] (ff3ast0) -- (ff3ast1);
    \draw[->] (ff0) -- (ff1);
    \draw[->] (ff1) -- (ff2);
    \draw[->, thick, orange] (ff2) -- node[left] {T} (ff3);
    \draw[->] (ff3) -- (ff4);
    \draw[->, thick, fuchsia] (ff2.west) -- +(-0.5, 0) node[left]{F} |- (ff4);
    \draw[->, dashed, thick, purple] (ff0.east) -- +(0.5, 0) |- (ff2);

    \end{tikzpicture}
    }
}
\caption{The comparison of the motivating example using the combination of CFG and AST. Black edges for sequential execute, orange edges for conditional true branch and fuchsia edges for the false branch. The dashed-purple edges are dataflow edges.}
\label{fig:the new program representation model}
\end{figure}

\section{Approach}
In this section, we would introduce our source code model and the learning model named MFGNN.

\subsection{Constructing Graph through Combining \zzhb{ECFG} and AST}
The combination of \zzhb{ECFG} and AST can be considered as a type of program dependency graph. The backbone of the graph is an inter-procedural CFG. A CFG $\mathcal{G}=(B, E)$ of the program is a directed graph, where $B$ is a collection of basic blocks and $E$ contains all control flow relationships. We choose three-address code (e.g., LLVM IR for C/C++ and Jimple for Java) as the intermediate representation when generating CFG by analysis frameworks (e.g., Clang for C/C++ and Soot for Java). %\yb{We applied the LLVM to parser C/C++ code. In LLVM, we used the SSA to obtain the conditional definition to a variable.}

From the motivating example (see Section~\ref{sec:1}), we can conclude that a precise modeling for basic block is essential for the following analysis. \zzhb{Therefore, we choose AST to model basic blocks as its nature of expressing syntactic structures and code information. 
To further enrich the information in the AST, we introduce the data types of variables to the subtrees of AST, which inherently lacks of such information and corresponds to variable usage (e.g., \texttt{DeclRefExpr} node in Clang).} Specifically, for the basic data type, we directly add it as a leaf node of the variable node. For user-defined classes, we refer to the method mentioned in~\cite{OpenVocabulary}, and separate these classes according to the Camel-Case naming. For type conversion statements, we process both the original type and the target type according to the above method and then add them into the AST as a subtree of the type conversion node. Another aspect we need to consider is the constants. To handle different constants, we disassemble the constant value bitwise~(e.g., The constant nodes 456 will be disassembled to three nodes, represent 4, 5, 6,respectively.).

To address the lack of dataflow dependency in both AST and CFG, in addition to the \textit{control flow}, \textit{call flow}, and \textit{exception flow} relationships included in the CFG, we also introduce \textit{data flow} relationship into our graph. Specifically, \textit{dataflow} relationship comes from the intra-procedural dataflow analysis. By traversing the CFG, we have built two collections of variables : $define$ stores variables defined in the block and $use$ stores variables used in the block. The dataflow relationship between basic blocks is obtained by reaching definition analysis later.
Then we divide the edges of \textit{control flow} into four categories according to their functionalities: \textit{sequential execution}, \textit{conditional true branch}, \textit{conditional false branch} and \textit{switch branches}. Different categories of flow relationship are labeled distinctly. Overall, there are seven types of edges in our source code model.

The final graph of the motivating example is shown in Figure~\ref{fig:the new program representation model}. We can observe that the red and green blocks of the two snippets respectively containing different AST, representing the great differences in \textit{local features} between blocks. \zzh{For the correct version (Fig. \ref{fig:1:b}), the AST of the green block indicates that a static field of that class is returned. For the faulty version (Fig. \ref{fig:1:a}), the AST of the red block indicates that a null value is returned. This slight textual difference, \textit{NO\_FIELDS} vs. \textit{null}, can be easily learnt with the help of a tree-based neural network due to the significant difference in the AST.} The control flow and the dataflow edges (i.e., dashed-purple edges) jointly describe the use of variables, and the conditional true edges (i.e., orange edges) indicate the branch and precondition of the faulty block. Combined with the difference in \textit{local features} between green and red blocks, our source code model leads to different \textit{context features}. 

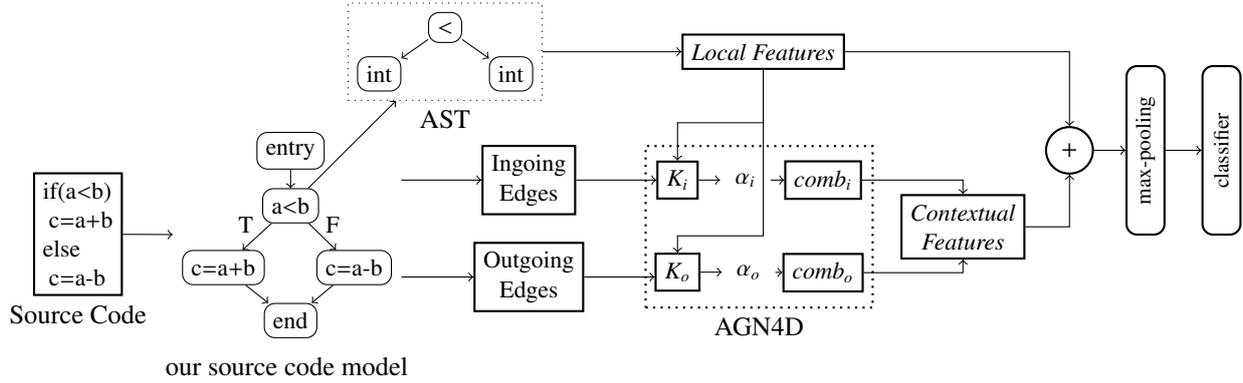
\begin{figure*}[!h]
\def\layerseq{2.5cm}
\centering
\begin{tikzpicture}
    [   node distance=\layerseq,
        ast_node/.style = {rectangle, rounded corners, draw, align=center, font=\small}]
    \node[rectangle, draw, thick, align=left, font=\small, label=below:Source Code] (code) {if(a$<$b)\\ \ c=a+b\\ else\\ \ c=a-b};
    \node[rectangle, rounded corners, draw, above right = 0.1em and 5em of code,align=left, font=\small] (cfg0) {entry};
    \node[rectangle, rounded corners, draw, below = 0.7em of cfg0, align=left, font=\small] (cfg1) {a$<$b};
    \node[rectangle, rounded corners, draw, below left = 1em and -0.1em of cfg1, align=left, font=\small] (cfg2) {c=a+b};
    \node[rectangle, rounded corners, draw, below right = 1em and -0.1em of cfg1, align=left, font=\small] (cfg3) {c=a-b};
    \node[rectangle, rounded corners, draw, below = 3em of cfg1, align=left, font=\small] (cfg4) {end};
    \node[label=below:our source code model, fit=(cfg0) (cfg1) (cfg2) (cfg3) (cfg4)] (cfg) {};

    \draw[->] (cfg0) -- (cfg1);
    \draw[->] (cfg1) -- node[above left=-0.3em of cfg2]{\small T} (cfg2);
    \draw[->] (cfg1) -- node[above right=-0.3em of cfg3]{\small F} (cfg3);
    \draw[->] (cfg2) -- (cfg4);
    \draw[->] (cfg3) -- (cfg4);
    \draw[->] (code) -- (cfg);

    \node[ast_node, above right = 3em and 1em of cfg] (ast0) {$<$};
    \node[ast_node, below right = 0.5em and 1em of ast0] (ast1) {int};
    \node[ast_node, below left = 0.5em and 1em of ast0] (ast3) {int};

    \draw[->] (ast0) -- (ast1);
    \draw[->] (ast0) -- (ast3);

    \node[draw, dotted, label=below:AST, fit=(ast0) (ast1) (ast3)] (ast) {};

    \draw[->] (cfg1) -- (ast);

    \node[draw, rectangle, thick, right =5.2em of ast, align=center, font=\small] (maps) {\textit{Local Features}};

    \node[draw, rectangle, thick, above right = -3.5em and 3em of cfg, align=center, font=\small] (inedge) {Ingoing\\Edges};
    \node[draw, rectangle, thick, below = 1em of inedge, align=center, font=\small] (outedge) {Outgoing\\Edges};

    \node[draw, rectangle, thick, right = 3em of inedge, align=center, font=\small] (gat00) {$K_i$};
    \node[draw, rectangle, thick, below = 2em of gat00, align=center, font=\small] (gat01) {$K_o$};

    \node[circle, right=0.8em of gat00, align=center, font=\small, minimum width=0.5em] (gat10) {$\alpha_i$};
    \node[circle, right=0.8em of gat01, align=center, font=\small, minimum width=0.5em] (gat11) {$\alpha_o$};

    \node[draw, rectangle, thick, right = 0.5em of gat10, align=center, font=\small] (gat20) {$comb_i$};
    \node[draw, rectangle, thick, below = 2em of gat20, align=center, font=\small] (gat21) {$comb_o$};

    \draw[->] (gat00) -- (gat10);
    \draw[->] (gat01) -- (gat11);
    \draw[->] (gat10) -- (gat20);
    \draw[->] (gat11) -- (gat21);

    \node[draw, dotted, rectangle, thick, label=below:AGN4D, fit=(gat00) (gat01) (gat10) (gat11) (gat20) (gat21)] (egat) {};

    \draw[<-] (inedge) -- (inedge-|cfg.east);
    \draw[<-] (outedge) -- (outedge-|cfg.east);
    \draw[->] (inedge) -- (inedge-|gat00.west);
    \draw[->] (outedge) -- (outedge-|gat01.west);
    \draw[->] (maps.south) -| +(0,-0.7) -| (gat00.north);
    \draw[->] (maps.south) -| +(0,-2.2) -| (gat01.north);

    \node[draw, rectangle, thick, align=center, right = 1em of egat, font=\small] (contextual) {\textit{Contextual}\\ \textit{Features}};

    \draw[->] (gat20.east) -| (contextual.north);
    \draw[->] (gat21.east) -| (contextual.south);
    \draw[->] (ast) -- (maps);

    \node[circle, draw, thick, above right = 1em and 1em of contextual] (add) {$+$};

    \draw[->] (contextual) -| (add);
    \draw[->] (maps) -| (add);

    \node[rectangle, rounded corners, draw, thick, right of=add, node distance=1cm, align=center, font=\footnotesize, minimum height=0.5cm, text width=2cm, rotate=90] (pool) {max-pooling};
    \node[rectangle, rounded corners, draw, thick, right of=pool, node distance=1cm, align=center, font=\footnotesize, minimum height=0.5cm, text width=2cm, rotate=90] (mlp) {classifier};

    \draw[->] (add) -- (pool);
    \draw[->] (pool) -- (mlp);
\end{tikzpicture}
\caption{MFGNN Structure}
\label{fig:model_structure}
\end{figure*}

\subsection{Multi-Flow Graph Neural Network}
We design a neural network model to obtain the features of our representation model and name it as Multi-Flow Graph Neural Network (MFGNN). Figure \ref{fig:model_structure} shows the overall structure of the model. The learning process can be divided into three stages: \textit{local features} embedding stage, \textit{contextual features} embedding stage and fusion stage. In the first stage, the tree-based network is used to learn the \textit{local features} for each block in $B$. In the second stage, attention-based graph neural network for directed graph (AGN4D) is used to learn \textit{contextual features} in the combined graphs based on \textit{local features} of each block.  In the final stage, a fusion layer is used to fuse \textit{local features} and \textit{contextual features}, and the contextual semantics are obtained.

\subsubsection{Local Features Embedding}
We use Tree-based Convolutional Neural Network (TBCNN) to obtain the \textit{local features} in each block. The original TBCNN is not suitable for our extended AST because of the additional contents on the leaf nodes of the extended AST.
The learning process of the original TBCNN ignores the fact that the deeper the node in the AST, the richer the information.
Therefore, we adjust the preset weights of TBCNN to increase the weight of deeper nodes in the convolution window of TBCNN, i.e., the nodes with richer information would have a more significant impact on the training process of the model. The formulas of the weights are shown as follows:
\begin{equation}
\eta^t_i = \frac{d_{i} - 1}{d_{max}-1}\ \ \eta^l_i = \eta^t_i \frac{p_i - 1}{n - 1}\ \ \eta^r_i = \eta^t_i (1 - \eta^l_i),
\end{equation}
, where $d_{i}$ is the depth of node $i$ in the entire tree, $d_{max}$ is the depth of entire tree, $p_i$ is the position of the node $i$ in subtree, and $n$ is the total number of $i$'s siblings.

The importance of \textit{local features} is two-fold: first, as the features of each block, \textit{local features} is the input for AGN4D (see Section~\ref{sec:context}) for learning \textit{contextual features}; second, \textit{local features} play an critical role within the final features, so we pass \textit{local features} to the fusion layer (see Section~\ref{sec:fusion}) directly.

\subsubsection{Contextual Features Embedding}\label{sec:context}
Our source code model can be considered as a directed graph with edge types. Therefore, based on GAT (see Section~\ref{sec:gat}), we design a network layer nested in MFGNN, named \textit{Attention-based Graph Neural Network for Directed Graph} (AGN4D), which can handle directed graph and multiple types of edges. With AGN4D, we can extract \textit{contextual features} from the combination graph.

Suppose $\mathcal{G}$ is an instance of the combined graph and $\mathcal{R}\mathcal{G}$ is the reverse graph of $\mathcal{G}$. Let $X=\{x_1,x_2,...,x_n\}$ represent \textit{local features} of the blocks in the set $B$ obtained in the previous stage.
Let the initial graph embedding $H^0 = \{h_1^0,h_2^0,...,h_n^0\}$ where $H^0 = X$, the graph embedding update process of $\mathcal{G}$ is as follows:
\begin{equation}
    \label{eqn:2}
    \begin{split}
    k^l_{o,u} &= MSG^l_o(h^{l-1}_u)\ \ \ \ k^l_{r,u} = MSG^l_r(h^{l-1}_u)\\
    h^l_{o,u} &= Agg_{o}^l(k^l_{o,u}, \{k^l_{o,v} | v \in \mathcal{N}_u^\mathcal{G}\})\ \ \ \ h^l_{r,u} = Agg_{r}^l(k^l_{r,u}, \{k^l_{r,v} | v \in \mathcal{N}_u^\mathcal{RG}\})\\
    h^l_u &= h^l_{o,u} + h^l_{r,u} + h^{l-1}_u
    \end{split}
\end{equation}
, where $\mathcal{N}_u^\mathcal{G}$ is the collection of successors of block $u$ in original graph $\mathcal{G}$ and $\mathcal{N}_u^\mathcal{RG}$ for reverse graph $\mathcal{RG}$. $MSG_o^l$ represents the $MSG$ function of the original graph at layer $l$ and $MSG_r^l$ for the reverse graph. $Agg_o^l$ refers to the $Agg$ function of the original graph at layer $l$ and $Agg_r^l$ for the reverse graph. Note that $MSG$ function and $Agg$ function do not share parameters between different layers. 

After obtaining the graph embedding from the two graphs, the graph embedding of previous layer and current layer are connected by a \textit{skip-connection} to obtain the final graph representation of this layer.

The $MSG$ function needs to transform the graph embedding from the previous layer to obtain the features of for this layer, which is parameterized by a weight matrix $W^l_{key}$ which is defined as follows:
\begin{equation}
    k^l_u = MSG^l(h^{l-1}_u) = W^l_{key} h^{l-1}_u.
\end{equation}

The $Agg$ function aggregates features in successor blocks and current block. 
\zzh{We add support for multiple types of edges to the \textit{self-attention} mechanism of GAT, defined as follows:}

\begin{equation}
        \begin{split}
        e_v^l &= P_{src}^l k_u^l + P_{dst}^{l,f} k_v^l\ \ \ <u,v,f>\ \in E\\
        \alpha_v^l &= softmax(\{LeakyReLU(e_v^l)|v \in \mathcal{N}_u\})\\
        h'_u &= Agg^l(k^l_u,\{k_v^l | v \in \mathcal{N}_u \}) = \sigma \left( \sum_{v \in \mathcal{N}_u}\alpha_{v}^lk_v^l \right)
        \end{split}
\end{equation}
, where $f$ stands for the flow type of the edge from $u$ to $v$. Attention mechanism is parameterized by $P_{src}^l$ and $P_{dst}^{l,f}$, which indicates the importance of the \textit{f}-type flow dependency between blocks \textit{u} and \textit{v}.

We pass the $h'_u$ of the last layer of AGN4D to the fusion layer as \textit{contexutal features}.

\subsubsection{Fusion Layer}\label{sec:fusion}
The main functionality of the fusion layer is to fuse \textit{local features} and \textit{contextual features} into the hybrid features of the program. In our design, the fusion layer first adds \textit{local features} and \textit{contextual features}, then gets the fixed size program feature vector through \textit{dynamic pooling}. In practice, we choose max-pooling as a pooling function. Finally, we train a classifier (i.e., Logistic Regression (LR)) for classification tasks.%to distinguish different categories.

\section{Evaluation}
We conducted a series of experiments to evaluate MFGNN with comparison against some existing state-of-art methods. Our experiments run on a 4 Tesla k40c GPUs machine with Xeon E5-2310 32GB RAM.

\subsection{Research Questions}
\zzh{
To evaluate the effectiveness of our source code model and MFGNN, and compare them with several state-of-the-art methods on some particularly tasks, our experiments were particularly designed to answer the following five research questions:
    \begin{itemize}
        \item [RQ1] How is the performance of MFGNN in classifying datasets that consists of programs with small textual but large semantic differences?
        \item [RQ2] How is the performance of MFGNN in Within-Project Defect Prediction (WPDP) task compared with the state-of-the-art methods?
        \item [RQ3] How is the performance of MFGNN in Cross-Project Defect Prediction (CPDP) task compared with the state-of-the-art methods?
        \item [RQ4] How is the performance of MFGNN in Functional Code-Clone Detection (CCD) task compared with the state-of-the-art methods?
        \item [RQ5] To what extent do different components in MFGNN influence the performance?
    \end{itemize}
}

%\subsection{Objects}
\subsection{Datasets}
\begin{table*}[!h]
\centering
  \caption{The statistics of program classification dataset for RQ1 and RQ5.}
  \label{tab:statistic}
  \begin{tabular}{c|cccc|ccccc}
    \toprule
    Index & \multicolumn{4}{c|}{CodeChef} & \multicolumn{5}{c}{Codeforces}\\
    \midrule
    {Problems} & \texttt{SUB} & \texttt{FLOW} & \texttt{MNMX} & \texttt{SUM} & \texttt{1062C} & \texttt{721C} & \texttt{731C} & \texttt{742C} & \texttt{822C} \\
    {Instance Count} & 2313 & 5487 & 9693 & 11666 & 9136 & 16084 & 10170 & 6971 & 17379\\
    {Avg. Line of Code} & 30 & 25 & 25 & 36 & 45 & 65 & 55 & 52 & 55\\
    {Avg. Branches Count} & 9 & 8 & 8 & 12 & 12 & 10 & 21 & 15 & 18 \\
    {Avg. Operators Count} & 25 & 15 & 15 & 35 & 40 & 40 & 29 & 30 & 39 \\
  \bottomrule
\end{tabular}
\end{table*}
\zzhb{
For RQ1 and RQ5, we selected two datasets as the objects of our experiments, namely CodeChef and Codeforces. The Codechef dataset is collected by Phan~\cite{url14} and composed of solutions, written in C/C++, which are submitted by users for four challenges, namely \texttt{SUB}, \texttt{MNMX}, \texttt{FLOW}, and \texttt{SUM}. 
However, these four challenges are trivial (e.g., \texttt{FLOW} only requires an implementation of the GCD algorithm), which cannot evaluate the effectiveness of our tool thoroughly. Thus, we further manually collected a dataset, namely Codeforces, from a public website\footnote{https://codeforces.com}. 
Specifically, it consists of solutions submitted by users for five challenges, i.e., \texttt{1062C}~\cite{1062C}, \texttt{721C}~\cite{721C}, \texttt{731C}~\cite{731C}, \texttt{742C}~\cite{742C} and \texttt{822C}~\cite{822C}. The challenges involved in the Codeforces dataset covers a variety of algorithms that are more complicated (e.g., disjoint-union sets, Dijkstra and greedy algorithm). Specifically, the detailed description of these challenges are described as follows:

\begin{itemize}
\item \textbf{\texttt{1062C}}: Given a binary-valued string and a list of intervals. , for each interval, the frequencies of each value in the interval is used to calculate a formula. A prefix sum (and product) algorithm is required to solve this challenge.
\item \textbf{\texttt{721C}}: Given a weighted directed graph, the shortest path is found between two specific nodes. Dijkstra algorithm is required to solve this challenge.
\item \textbf{\texttt{731C}}: Given an undirected graph, the number of connected components in the graph is counted. A disjoint-union sets is required to solve this challenge.
\item \textbf{\texttt{742C}}: Given a directed graph, the least common multiplier (LCM) is calculated for the lengths of all the circles in the graph. To solve this challenge correctly, circle finding algorithm and LCM algorithm are required.
\item \textbf{\texttt{822C}}: Given a collection of weighted intervals, a subset of the minimum weight sum is found to satisfy some conditions (e.g., no intersect between intervals). A greedy algorithm is required to solve this challenge.
\end{itemize}
}

For each program in both datasets, there is a label to indicate the running result of the corresponding program. The meaning of labels is detailed in the following: 

\begin{itemize}
\item \textbf{Accepted (AC)}: The program is able to pass all test cases; 
\item \textbf{Wrong Answer (WA)}: The program can execute normally but output incorrect results; 
\item \textbf{Runtime Error (RE)}: The program cannot execute normally on some test cases, which are generally due to illegal memory access or operation error, e.g., divided by zero; 
\item \textbf{Time Limited Exceeded (TLE)}: The program does not response within the time limits;
\item \textbf{Memory Limited Exceeded (MLE)}: The consumed resource, i.e., memory, exceed the requirement.
\end{itemize}

\zzhb{
Except for the \textbf{AC}, different running results correspond to different defects in source code. For example, the source code with the \textbf{TLE} often contains redundant steps or dead loops, while the source code with the \textbf{WA} often contains functional errors. Therefore, we argue that a reasonable source code model should reflect these differences and is able to classify them effectively.

Additionally, we conducted a pre-processing on both datasets.
First, we removed the source code that are irrelevant to the corresponding challenge.
Second, we removed the duplicated ones from datasets.
Third, to avoid mislabeling, we generated some test cases according to the requirements of the corresponding challenge. Then, we re-ran the source code and re-labeled them that were mis-labeled.
Finally, for each dataset of challenges, we split each of them into training set, validation set and test set in 3:1:1 ratio. Table~\ref{tab:statistic} shows some metrics of the final datasets.}

For RQ2 and RQ3, we have selected another well-known public dataset, namely PROMISE. The reason is that it has been widely used for software defect prediction~\cite{url1,url12, DTLDP}, and it consists of several well-known open-source Java projects.
Except for the jedit (Version 3.2), which cannot be compiled properly, the remaining 10 Java projects and their corresponding versions that we selected are identical to a previous work~\cite{url1} for comparison.
Finally, 1395 source code files, which cannot be processed successfully by our Soot-based generator, were removed from the dataset.
The statistical description of the final dataset for RQ2 and RQ3 is shown in Table~\ref{tab:promise_stat}.

 \begin{table}[!h]
 \footnotesize
    \caption{The statistics of PROMISE dataset, which is specialized for RQ2 and RQ3}
    \label{tab:promise_stat}
    \centering
    \begin{tabular}{ccccc}
    \toprule
    App & Ver & Mean files & Mean defective & Defective rate \\
    \midrule
    lucene & 3 & 247 & 140 & 56.7 \\
    synapse & 3 & 188 & 52 & 27.7 \\
    xerces & 2 & 295 & 54 & 18.3 \\
    xalan & 2 & 665 & 237 & 35.6 \\
    camel & 3 & 700 & 165 & 23.6 \\
    log4j & 2 & 70 & 29 & 41.4 \\
    ant & 3 & 422 & 95 & 22.5 \\
    jedit & 3 & 311 & 67 & 21.5 \\
    poi & 3 & 328 & 219 & 66.8 \\
    ivy & 2 & 253 & 26 & 10.3 \\
    \bottomrule
    \end{tabular}
\end{table}
 
\zzhb{
For the remaining research question, i.e., RQ4, we have selected a public dataset, namely OJClone, which has been adopted by several works~\cite{url13, FCD}. It was collected from an online program judgement system for C/C++ source code.
Specifically, OJClone contains 15 program tasks, and each of them is composed of 500 source code files submitted by users.
For the same task, different users' source codes could pass the test and got \textbf{AC} verdict, and thus can be considered as functional code clone. In other words, for each source code pair in the dataset, it will be labeled by either 0 for non-cloned pair or 1 for cloned pair.
Similarly to the classifying task, we shuffled and split the dataset into training, validation and testing in 3:1:1 ratio.}

%\subsection{Variables}
\subsection{Experiment Settings}
In this section, we present the setup of each RQ's experiment, involving detailed settings about our method, the choices of baseline methods and comparison metrics.

\subsubsection{Settings for MFGNN}
The input of MFGNN consists of four parts: 1) a collection of AST nodes (represented by one-hot vectors); 2) a collection of AST's substructures; 3) a mapping graph (i.e., mapping the substructure to corresponding basic block); and 4) an \zzhb{ECFG} 
As for the hyper-parameters, the embedding dimension of the AST nodes is set as 50. And the dimension of AGN4D, which is stacked with three layers, was set as 200. MFGNN was optimized by Adamax, and trained for 200 epochs. During the training, we selected the parameters~(i.e., weights of MFGNN) that performed best on the validation set, and evaluated them on the test set.

\subsubsection{Settings for Baselines}
\paragraph{For RQ1} To illustrate the effectiveness of MFGNN, we choose three other well-known groups of representative methods for comparison:

\textbf{SVM-based approaches} We chose SVM-based approaches to demonstrate that both datasets, i.e., CodeChef and Codeforces, do consist of source code with small textual but large semantic distinctions. 
In terms of classifying source code files according to their textual features, the more indistinguishable the source code are, the worse SVM-based methods would perform.
To show the textual distinguishability of our dataset, we choose TF-IDF and BoW features as the textual features, and feed them into RBF-kernel SVM.

\textbf{AST-based approaches} To illustrate the advantages of our source code model over AST in program classification, we chose several typical AST-based approaches. Specifically, according to AST granularity, we can divide the AST-based approaches into two categories. One uses the entire AST of source code, like representative methods: TBCNN~\cite{url10} and Tree-LSTM~\cite{url44}. The other one splits AST according to code fragments and is known as ASTNN~\cite{url13}. \zzhb{Moreover, code2vec~\cite{c2v} adopts paths in AST to represent the source code and learns the features contained in the paths through a network based on attention mechanisms. Similarly, code2seq~\cite{c2s} uses the same paths as code2vec but extracts the features by the seq2seq model~\cite{seq2seq}.}

For the settings of AST-based approaches, the AST used in TreeLSTM, TBCNN and ASTNN is generated by Clang, but the AST paths used by code2vec and code2seq are generated by ASTMiner\footnote{https://github.com/JetBrains-Research/astminer}.
For code2vec, the embedding dimension is set to 400; For code2seq, the embedding dimension is set to 128 and the decoder dimension is set to 320; The hidden dimension of the other methods is set to 200.

\textbf{Graph-based approaches} Some recent studies focused on representing a program as a graph and adopting a graph-based learning method to extract dependency features from the graph. 
DGCNN chooses CFG as the source code model and obtains features with GCN~\cite{url14}. ContextGraph (CtxG) inserts extra edges (e.g., dataflow edges) into the original AST, and extracts the features with GGNN~\cite{url15}. 
For the settings of graph-based approaches, the number of steps of GGNN is set to 3, and the hidden size of all graph-based approaches was set to 200.

%\he{The paragraphs corresponding to RQ2 and RQ3 are too tedious, try your best to revise.}

\paragraph{For RQ2} We evaluated the performance of MFGNN on Within-Project Defect Prediction (WPDP) task.
According to previous studies on defect prediction task~\cite{url1,url12}, we decided to use the same strategy, i.e., training by the earlier version and predicting on the later version. We compared MFGNN with several typical WPDP methods that can be divided into two types according to their adopted source code models.
Some of defect prediction technologies used the features of the PROMISE with traditional machine learning methods~\cite{ml-1, ml-2}, including Adaboost, Multi-Layer Perception (MLP) and Random Forest (RF). The others utilize AST-based features, and representative methods (e.g., DBN~\cite{url1} and TreeLSTM~\cite{url12}). 
Specifically, DBN obtained the semantic features from AST. We classified these features with three classifiers: Naive-Bayes ($\textrm{DBN}_{NB}$), Logistic Regression ($\textrm{DBN}_{LR}$) and Decision Tree ($\textrm{DBN}_{DT}$). As for TreeLSTM, after the AST was parsed by JavaParser\footnote{https://javaparser.org}, it would take the entire AST as input for prediction.
Additionally, we chose another two well-known methods : DTL-DP~\cite{DTLDP} and BugContext~\cite{Li2019ImprovingBD}. The former one visualized the source code file (or binary file) as an image, and obtained the defect features with AlexNet, while the later one acquired contextual dependencies from CFG and DFG, then introduced them into path-based AST features.

%As for the source code model, the one for DBN is semantic features that is obtained from AST, and the one for TreeLSTM is the entire AST, which is parsed by JavaParser\footnote{https://javaparser.org}. For DBN, we chose three ways to classify the semantic features: .For the methods based on other source code models (e.g., visualization), we chose two well-known methods: DTLDP~\cite{DTLDP} and BugContext~\cite{Li2019ImprovingBD}. 

\paragraph{For RQ3} We conducted Cross-Project Defect Prediction~(CPDP) experiments to show the performance of MFGNN. Following the previous studies~\cite{url1,url12}, we organized ten groups of experiments, trained models on the \textit{source project} and predicted on the \textit{target project}.
For the target project, according to transfer learning methods~\cite{url22}, we first randomly selected 30\% of the data to fine-tune a LR-based classifier and then predicted the rest 70\%. Except for DBN, which was replaced by its CPDP-variant: DBN-CP~\cite{url1}, we chose the same set of baseline methods as RQ2. Additionally, we added two transfer learning-based methods, namely TCA+~\cite{url22} and TNB~\cite{Maying2012TransferLF}, which take the PROMISE feature as same as the machine learning methods.

%We chose DBN-CP, the CPDP variant of DBN, for comparison. Except for DBN, based on the baselines mentioned in RQ2 part, we added two transfer learning-based methods additionally, namely TCA+~\cite{url22} and TNB~\cite{Maying2012TransferLF}, which take the PROMISE feature as same as the machine learning methods.

\paragraph{For RQ4} We conducted Functional Code-Clone Detection~(CCD) experiments to demonstrate the distinguishability of semantics obtained by MFGNN. Let the features of the two source code files within a pair that are obtained from MFGNN be $v_1$ and $v_2$, respectively. The difference can be defined as $d = |v_1 - v_2|$. Finally, we use a LR-based classifier~(i.e., $y = sigmoid(W_od+b_o)$) to determine whether the code pairs are similar based on the vector $d$. We compared the performance of MFGNN with several state-of-the-art models that are widely used on CCD task, including RAE+~\cite{RAE}, Deckard~\cite{Deckard}, CDLH~\cite{CDLH}, ASTNN~\cite{url13}, DeepSim~\cite{deepsim}, and FCDetect~\cite{FCD}.

\paragraph{For RQ5} We carried out some ablation studies. \zzh{Our approach can be divided into two parts, a source code model based on \zzhb{ECFG} and a learning model with the AGN4D layer. 
\zzhb{Firstly, we explored the impact of different choices in the design of our source code model, which has four options: 1) representing basic blocks with AST (\textbf{A}) or BoW features (\textbf{B}); 2) including control flow edges (\textbf{C}) or not; 3) including dataflow edges (\textbf{D}) or not; and 4) embedding the source code model with multi-typed edge~(\textbf{M}) or with single-typed edge~(\textbf{S}).
We have designed four variants based on the combination of different options.

\begin{itemize}
    \item \textbf{A}ST+\textbf{C}FG+\textbf{S}ingle: The main body of this model is CFG with no distinction between control flow types, and its basic blocks are represented using ASTs.
    \item \textbf{A}ST+\textbf{D}FG+\textbf{S}ingle: The main body of this model is DFG, with only one type of flow, and its basic blocks are represented using ASTs.
    \item \textbf{A}ST+\textbf{C}FG+\textbf{M}ulti: The main body of this model is a CFG that distinguishes between different control flows, and its basic blocks are represented using ASTs.
    \item \textbf{B}oW+\textbf{C}FG+\textbf{D}FG+\textbf{M}ulti: The main body of this model is a CFG that contains the dataflows and distinguishes between different types of flows. Its basic blocks are represented using BoW.
\end{itemize}}
Secondly, we explored the impact of different graph learning methods.}
We replaced the AGN4D layer with graph convolution network (\textbf{GCN}) and gated-graph neural network (\textbf{GGNN}), respectively. \zzhb{Additionally, we compared across the different options in AGN4D, i.e., summation and concatenation, to synthesize graph features (see Eq.\ref{eqn:2}) on the same source code model.}

%\subsubsection{Dependent Variables}
\subsubsection{Metrics}
For RQ1 and RQ5, we chose the \textit{accuracy} and \textit{macro-F1}~\cite{macrof1} to evaluate the prediction result on test sets. Assuming a task has $K$ classes, the \textit{accuracy} is defined as follow:
\begin{equation}
accuracy=\frac{\sum_{i=1}^{K}TP_i}{N},
\end{equation}
, where $TP_i$ refers to true positive of class $i$, and $N$ is the total number of samples.

For a binary classification task, the \textit{F1-score}~(F1) is defined as follow:
% measures the weighted harmonic mean of the \textit{precision} and \textit{recall}, where the \textit{precision} refers to the ratio of correctly predicted defective files over all the files predicted as being defective, and the \textit{recall} refers to the ratio of correctly predicted defective files over all of  the true defective files. The formula for calculating \textit{F1-score} is as follows:
\begin{equation}
F1\textrm{-}score=\frac{2*precision*recall}{precision+recall}
\end{equation}
, where $precision=\frac{TP}{TP+FP}$ and $recall=\frac{TP}{TP+FN}$, $TP$ denotes the true positive, $FP$ represents the false positive, and $FN$ refers to false negative.

A multi-label classification task can be considered as several binary classification tasks on different labels. Based on that, assuming the task has $K$ classes, the \textit{macro-F1} can be defined as follow:
\begin{equation}
Macro\textrm{-}F1=\frac{1}{K}\sum_{i=1}^{K}F1\textrm{-}score_i.
\end{equation}

For RQ2 and RQ3, in addition to the F1 on the buggy class, we also used the metric \textit{AUC} (Area Under the receiver operating characteristics Curve)~\cite{url12} to evaluate the performance of defect prediction. Specifically, \textit{AUC} refers to the probability of a classifier ranking a randomly selected positive sample higher than a randomly selected negative sample. Intuitively speaking, a higher value of \textit{AUC} implies a better performance.

\zzh{For RQ4, following the evaluation metrics of previous works~\cite{url13, FCD}, we choose \textit{precision} (P), \textit{recall} (R) and F1 to measure the performance of the selected models on CCD task.}

\section{Results}
In this section, we show the results of the experiments, and compare the performance of different methods.
\begin{table*}[!h]
\centering
\setlength\tabcolsep{1.8pt}
%\small
\caption{Results on program classification task, the numbers in parentheses are the parameter sizes of methods.}
\label{table:program_classification}
\resizebox{\textwidth}{!}{
\begin{tabular}{cccccccccccccccccccc|cc}
    \toprule
    \multirow{2}{*}{Groups} & \multirow{2}{*}{Methods} & \multicolumn{2}{c}{SUB} & \multicolumn{2}{c}{MNMX} & \multicolumn{2}{c}{FLOW} & \multicolumn{2}{c}{SUM} & \multicolumn{2}{c}{1062C} & \multicolumn{2}{c}{721C} & \multicolumn{2}{c}{731C} & \multicolumn{2}{c}{742C} & \multicolumn{2}{c|}{822C} & \multicolumn{2}{c}{Avg}\\
    & & Acc & F1 & Acc & F1 & Acc & F1 & Acc & F1 & Acc & F1 & Acc & F1 & Acc & F1 & Acc & F1 & Acc & F1 & Acc & F1\\
    \midrule
    \multirow{2}{*}{SVM} 
    &SVM\&TF-IDF&34.7 & 12.9 & 48.0 & 16.2 & 56.6 & 18.1 & 38.4 & 13.9 & 51.0 & 13.6 & 38.7 & 11.2 & 42.9 & 12.0 & 60.0 & 15.0 & 56.2 & 14.4 & 47.4 & 14.1\\
    &SVM\&BoW&54.5 & 41.5 & 68.6 & 52.5 & 80.0 & 60.8 & 59.9 & 52.6 & 55.7 & 21.5 & 42.6 & 22.6 & 55.4 & 33.1 & 66.9 & 26.2 & 56.8 & 16.6 & 60.0 & 36.4 \\
    \midrule
    \multirow{5}{*}{AST} 
    &TBCNN (0.5M)&67.2 & 65.2 & 74.6 & 69.2 & 75.3 & 66.0 & 63.8 & 62.4 & 63.0 & 39.9 & 53.7 & 47.1 & 65.6 & 52.9 & 66.9 & 38.8 & 58.9 & 48.9 & 65.4 & 54.5\\
    &TreeLSTM (4.0M)&66.1 & 64.1 & 76.0 & 69.5 & 76.8 & 68.4 & 66.3 & 65.9 & 66.9 & 47.7 & 56.0 & 50.6 & 69.1 & 53.1 & 70.0 & 41.2 & \textbf{60.7} & 50.2 & 67.5 & 56.7\\
    &ASTNN (0.9M)&61.4 & 58.9 & 70.3 & 63.1 & 74.3 & 62.4 & 62.7 & 62.4 & 63.6 & 46.6 & 49.9 & 44.3 & 61.2 & 50.0 & 64.6 & 32.6 & 55.0 & 42.3 & 62.6 & 51.4\\
    &code2vec (173M)&29.5&24.7&36.6&25.7&29.7&21.1&31.4&24.5&41.2&18.4&28.9&18.5&28.1&18.2&49.2&18.0&30.7&14.5&33.9&20.4\\
    &code2seq (61M)&35.9&16.9&50.4&16.9&51.7&30.0&31.9&21.3&51.4&13.6&36.0&15.1&43.0&13.9&52.3&17.3&56.5&14.5&45.5&17.7\\
    \midrule
    \multirow{3}{*}{Graph} &DGCNN (0.4M)&64.8 & 64.5 & 74.6 & 67.7 & \textbf{83.8} & 70.9 & 69.1 & 67.4 & 64.3 & 42.8 & 54.2 & 49.6 & 61.4 & 47.0 & 70.3 & 44.1 & 56.2 & 44.5 & 66.5 & 55.4\\
    &DGCNN (2.4M)&64.4&62.6&74.2&66.5&82.7&\textbf{72.0}&69.1&67.9&64.8&42.1&55.3&49.9&61.7&48.5&72.6&43.5&55.9&42.5&66.7&55.1\\
    &CtxG (4.9M)&64.8 & 62.0 & 74.0 & 68.0 & 74.9 & 63.9 & 64.9 & 64.6 & 59.1 & 42.0 & 51.1 & 45.3 & 59.0 & 47.8 & 65.0 & 36.8 & 56.4 & 43.3 & 63.2 & 52.6\\
    \midrule
    \multicolumn{2}{c}{\textbf{MFGNN (2.1M)}} & \textbf{74.5} & \textbf{74.7} & \textbf{83.1} & \textbf{81.4} & 81.8 & 71.0 & \textbf{72.9} & \textbf{73.5} & \textbf{68.0} & \textbf{53.2} & \textbf{59.5} & \textbf{54.5} & \textbf{70.0} & \textbf{61.0} & \textbf{73.8} & \textbf{51.6} & 59.9 & \textbf{50.3} & \textbf{71.5} & \textbf{63.5}\\
    \bottomrule
\end{tabular}
}
\end{table*}

\subsection{Answer to RQ1}
Table~\ref{table:program_classification} illustrates the results related to RQ1, and the best performance are highlighted in bold. In column 2, we list the size of the corresponding model except for the SVM-based approaches, whose size is neglectable. According to these experimental results, we have the following insights:

\zzh{\textbf{\textit{The dataset does consist of source code with a minimal textual difference.}}}
As we can see, SVM-based methods did not play well in our experiments, which is reflected by their corresponding F1 values. This indicates that the source codes with different labels in our dataset cannot be effectively distinguished by textual features. In other words, it proves that the textual differences among the source codes in out dataset are too small to be distinguished effectively.

\zzhb{\textbf{\textit{Compared to AST-based approaches, MFGNN achieves a better performance with fewer parameters.}}
Compared to the best method, i.e., TreeLSTM, among AST-based approaches, MFGNN reduces the model parameters by up to 50\%, while achieving 4.0\% and 6.8\% improvements on accuracy and F1, respectively.
Additionally, we can observe that both of code2vec and code2seq did not perform well. This is because both of them model the source code by sampling the path of the AST, which can only capture potential connections between code tokens~\cite{c2ve}. Program classification task, however, requires the identification of the actual control flow and dataflow information of the program execution, which can not be achieved by their models.
On the contrary, our source code model can reflect the actual execution path of the program with contextual information, which can be better captured by the neural network.}

\textbf{\textit{MFGNN achieves a significant performance improvement while adding a limited number of parameters compared with the graph-based approaches.}} Compared to the best graph-based approach, DGCNN, MFGNN only increases the number of parameters by 4 times, but achieves  5\% and 8.1\% improvement on accuracy and F1, respectively. Similarly, compared with DGCNN, which has the same scale of parameters as MFGNN, MFGNN achieves 4.8\% and 8.4\% improvement on accuracy and F1, respectively. 
This result illustrates that the performance of MFGNN has little correlation with its number of parameters. The main difference between MFGNN and traditional graph-based methods is two-fold. On one hand, the integration of multiple flow information in the source code model clearly expresses the dependency features of the program well.
On the other hand, the attention mechanism allows MFGNN to dynamically adjust the weights of different types of flows, \zzhb{resulting in a better mining of the flow features.}

\begin{table*}[!h]
    \setlength\tabcolsep{1.6pt}
%    \small
    \centering
    \caption{The result of WPDP experiment on PROMISE.}
    \label{table:promise_wpdp}
    \resizebox{\textwidth}{!}{
    \begin{tabular}{ccc|cc|cc|cc|cc|cc|cc|cc|cc|cc|cc}
        \toprule
        \multicolumn{3}{c}{Methods} & \multicolumn{2}{c}{Adaboost} &\multicolumn{2}{c}{MLP} & \multicolumn{2}{c}{RF} & \multicolumn{2}{c}{$\textrm{DBN}_{NB}$} &  \multicolumn{2}{c}{$\textrm{DBN}_{LR}$} &  \multicolumn{2}{c}{$\textrm{DBN}_{DT}$} & \multicolumn{2}{c}{Tree-LSTM} & \multicolumn{2}{c}{DTLDP} &\multicolumn{2}{c}{BugContext} &\multicolumn{2}{c}{MFGNN}\\
        \midrule
        Project & Tr & T & F1 & AUC & F1 & AUC & F1 & AUC &F1 & AUC & F1 & AUC & F1 & AUC & F1 & AUC & F1 & AUC & F1 & AUC & F1 & AUC\\
        \midrule 
        \multirow{2}{*}{ant} & 1.5 & 1.6 & 37.8 & 68.4 & 32.0 & 72.5 & 36.2 & 70.5 & 4.3 & \textbf{81.5} & 40.7 & 80.7 & 4.3 & 51.1 & 29.7 & 49.7 & \textbf{45.3} & 22.8 & 31.1 & 44.4 & 33.1 & 72.5\\
 & 1.6 & 1.7 & 52.2 & 69.4 & 51.4 & 71.6 & 49.1 & 74.1 & 53.2 & 69.3 & 51.7 & \textbf{79.0} & 22.8 & 50.6 & 44.2 & 60.8 & 35.5 & 50.0 & 45.1 & 44.7 & \textbf{53.7} & 75.5\\

\midrule
\multirow{2}{*}{camel} & 1.2 & 1.4 & 40.2 & 70.3 & 39.8 & 68.6 & 47.2 & 75.9 & 12.9 & 53.1 & 16.5 & 40.4 & 9.3 & 51.9 & 53.1 & 82.7 & 32.9 & 34.1 & 36.2 & 52.6 & \textbf{54.3} & \textbf{83.6}\\
 & 1.4 & 1.6 & 40.2 & 70.9 & 30.7 & 68.9 & 45.9 & 70.1 & 13.7 & 58.4 & 32.0 & 58.4 & 8.0 & 44.2 & 55.9 & 79.7 & 34.7 & 44.2 & 27.8 & 50.3 & \textbf{56.8} & \textbf{84.0}\\

\midrule
\multirow{1}{*}{ivy} & 1.4 & 2.0 & 14.3 & 66.9 & 14.8 & 67.8 & 23.1 & \textbf{69.4} & \textbf{47.6} & 61.5 & 27.3 & 57.9 & 26.7 & 57.7 & 15.9 & 45.8 & 21.1 & 18.5 & 31.9 & 44.5 & 22.9 & 60.2\\

\midrule
\multirow{1}{*}{jedit} & 4.0 & 4.1 & 57.0 & 80.7 & 54.3 & 80.4 & 54.5 & 79.9 & 41.3 & 45.6 & 41.6 & 50.0 & 0.0 & 50.4 & 62.0 & 78.8 & 23.8 & 35.7 & 38.5 & 63.1 & \textbf{65.0} & \textbf{84.4}\\

\midrule
\multirow{2}{*}{lucene} & 2.0 & 2.2 & 58.5 & 63.7 & 59.9 & 63.4 & 59.4 & \textbf{65.7} & 32.7 & 65.3 & 36.6 & 65.4 & 35.8 & 53.3 & 60.9 & 59.9 & 58.9 & 48.0 & 43.0 & 58.4 & \textbf{64.6} & 64.0\\
 & 2.2 & 2.4 & 64.8 & 56.6 & 68.4 & 57.5 & 64.8 & 62.1 & 25.7 & 47.3 & 37.4 & \textbf{73.3} & 14.2 & 71.6 & 68.1 & 59.1 & \textbf{68.8} & 40.3 & 68.0 & 60.3 & \textbf{68.8} & 63.4\\

\midrule
\multirow{1}{*}{log4j} & 1.0 & 1.1 & 66.7 & 78.0 & 73.3 & 82.5 & 75.0 & 84.2 & 75.0 & 88.5 & 60.5 & \textbf{90.2} & 72.3 & 64.8 & 73.3 & 75.8 & 24.0 & 46.9 & \textbf{75.5} & 66.7 & 73.3 & 77.0\\

\midrule
\multirow{2}{*}{poi} & 1.5 & 2.5 & 77.3 & 72.6 & 78.4 & 72.1 & 73.3 & 74.3 & 8.5 & 45.8 & 8.4 & 65.4 & 13.4 & 40.9 & 81.6 & 75.8 & 81.9 & 59.5 & 79.7 & 62.1 & \textbf{83.1} & \textbf{78.4}\\
 & 2.5 & 3.0 & 54.6 & 50.2 & 68.4 & 52.2 & 58.7 & 55.6 & 28.0 & 76.4 & 27.0 & 78.6 & 8.9 & \textbf{78.7} & 73.9 & 69.5 & \textbf{77.7} & 71.9 & 65.2 & 58.3 & 73.3 & 69.2\\

\midrule
\multirow{2}{*}{synapse} & 1.0 & 1.1 & 28.9 & 64.6 & 15.0 & 61.1 & 14.7 & 57.9 & 47.9 & 64.4 & 43.0 & \textbf{66.3} & \textbf{48.9} & 60.5 & 28.2 & 43.2 & 41.0 & 51.7 & 18.8 & 40.1 & 30.4 & 61.1\\
 & 1.1 & 1.2 & 40.3 & 61.2 & 44.1 & 64.4 & 40.0 & 66.8 & 41.5 & \textbf{69.1} & 41.5 & 50.1 & 35.9 & 66.5 & 50.3 & 57.8 & \textbf{54.4} & 43.7 & 42.4 & 55.0 & 50.3 & 65.6\\

\midrule
\multirow{1}{*}{xalan} & 2.4 & 2.5 & 32.9 & 62.1 & 21.9 & 59.7 & 27.9 & 59.1 & 19.1 & 51.1 & 30.8 & 58.2 & 10.6 & 55.4 & 34.5 & \textbf{63.9} & \textbf{50.4} & 43.8 & 17.4 & 51.9 & 33.1 & 58.7\\

\midrule
\multirow{1}{*}{xerces} & 1.2 & 1.3 & 29.6 & 62.6 & 24.2 & 60.3 & 25.7 & 57.9 & 24.1 & 53.5 & 32.4 & 64.0 & \textbf{33.3} & 64.5 & 29.4 & 60.7 & 14.8 & 29.4 & 9.4 & 51.5 & 30.9 & \textbf{74.2}\\

\midrule \multicolumn{3}{c|}{Avg} & 46.4 & 66.5 & 45.1 & 66.9 & 46.4 & 68.2 & 31.7 & 62.1 & 35.2 & 65.2 & 23.0 & 57.5 & 50.7 & 64.2 & 44.3 & 42.7 & 42.0 & 53.6 & \textbf{52.9} & \textbf{71.5}\\
        \bottomrule 
    \end{tabular}
    }
\end{table*}

\subsection{Answer to RQ2}\label{sec:answer-rq2}
Table~\ref{table:promise_wpdp} shows the performance of different approaches on the within-project defect prediction (WPDP) task, and the best performances are highlighted in bold. Due to the limitations of Soot~(e.g., throw exceptions on some data items), our dataset lost a large number of entries in some projects, which resulted in the distribution of the dataset we actually used differs from the previous study~\cite{url1}. To ensure the fairness of the comparison, we re-implemented the DBN methods and TreeLSTM mentioned in~\cite{url12}. We selected multiple groups of parameters randomly, ran all methods multiple times and kept the best result.

Compared with the state-of-art method, namely TreeLSTM, MFGNN achieved 1.6\% and 4.0\% improvements on F1 and AUC, respectively. Moreover, MFGNN was 5\% and 29.6\% higher in F1 and AUC, respectively, than DTLDP. Specifically, higher AUC often means that the model has more confidence in the prediction results, and the main difference between MFGNN and these methods is the use of \bo{ECFG} on the source code model allows MFGNN to capture contextual dependencies.

Compared to the BugContext method, MFGNN improved 7.3\% and 18.7\% in F1 and AUC, respectively. We think such a significant improvement can be attributed to their structural difference, which can be divided into three-fold. First, \textit{\textbf{the representation of basic blocks.}} According to the open-source implementation of the BugContext, it only embeds line numbers into basic blocks, while MFGNN uses AST to represent those basic blocks. Second, \textit{\textbf{the process of learning AST features.}} BugContext learns tree features by sampling the paths of the tree, while TBCNN is adopted to learn the features by MFGNN. Third, \textit{\textbf{the process of learning graph.}} MFGNN uses AGN4D to capture the dependency features in the graph, while BugContext uses node2vec to learn the information in the PDG. The biggest advantage of AGN4D over node2vec is the introduction of an attention mechanism, which allows different types of dependency features to be \bo{fused}. \zzhb{In conclusion, the hybrid features obtained by MFGNN could perform better on WPDP task.}

\begin{table*}[!h]
    \centering
\setlength\tabcolsep{2pt}
%    \small
    \caption{The result of CPDP experiment on PROMISE}
    \label{table:promise_cpdp}
    \resizebox{\textwidth}{!}{
    \begin{tabular}{c|c|cc|cc|cc|cc|cc|cc|cc|cc|cc}
    \toprule
    \multirow{2}{*}{Source} & \multirow{2}{*}{Target} & \multicolumn{2}{|c|}{Adaboost} & \multicolumn{2}{|c|}{MLP} & \multicolumn{2}{|c|}{RF}& \multicolumn{2}{|c|}{TCA+} & \multicolumn{2}{|c|}{TNB} & \multicolumn{2}{|c|}{DBN-CP} & \multicolumn{2}{|c|}{DTLDP} & \multicolumn{2}{|c|}{BugContext} & \multicolumn{2}{|c}{MFGNN}\\
    & & F1 & AUC & F1 & AUC & F1 & AUC & F1 & AUC & F1 & AUC & F1 & AUC & F1 & AUC & F1 & AUC & F1 & AUC\\
    \midrule
    ant-1.6 & camel-1.4  & 23.9 & 62.9 & 37.2 & 63.3 & 26.8 & 65.9 & 28.0 & 52.9 & \cellcolor{blue!25}\textbf{40.0} & \cellcolor{blue!25}\textbf{67.5} & 31.9 & 60.7 & 22.8 & 42.0 & 22.6 & 42.5 & \cellcolor{red!25}36.3 & \cellcolor{red!25}65.3\\
jedit-4.1 & camel-1.4  & 25.7 & 60.4 & 26.5 & 49.4 & 16.7 & 59.7 & 29.0 & 51.2 & \cellcolor{blue!25}32.0 & \cellcolor{blue!25}64.2 & 23.4 & 61.1 & 31.3 & 54.9 & 11.7 & 51.6 & \cellcolor{red!25}\textbf{39.8} & \cellcolor{red!25}\textbf{67.4}\\
\midrule
camel-1.4 & ant-1.6  & 54.3 & 69.4 & 32.8 & 38.1 & 38.2 & 70.1 & 25.0 & 42.1 & \cellcolor{blue!25}\textbf{59.0} & \cellcolor{blue!25}\textbf{79.0} & \cellcolor{red!25}56.1 & \cellcolor{red!25}74.3 & 47.8 & 18.6 & 22.2 & 66.9 & 50.3 & 71.4\\
poi-3.0 & ant-1.6  & 48.8 & 68.7 & \cellcolor{blue!25}\textbf{62.7} & \cellcolor{blue!25}\textbf{80.3} & 55.0 & 73.8 & 28.0 & 46.8 & 53.0 & 73.6 & 48.2 & 63.3 & 44.8 & 26.5 & 51.0 & 68.2 & \cellcolor{red!25}56.9 & \cellcolor{red!25}75.2\\
\midrule
camel-1.4 & jedit-4.1  & 34.8 & 57.7 & 13.2 & 30.2 & 37.6 & 71.1 & 50.0 & 63.8 & \cellcolor{blue!25}\textbf{53.0} & \cellcolor{blue!25}\textbf{76.2} & 32.3 & 59.1 & 38.4 & 36.7 & \cellcolor{red!25}45.2 & \cellcolor{red!25}70.2 & 41.5 & 64.2\\
log4j-1.1 & jedit-4.1  & 57.7 & 78.6 & 56.1 & 72.1 & 56.6 & 78.5 & 18.0 & 35.7 & \cellcolor{blue!25}\textbf{62.0} & \cellcolor{blue!25}\textbf{79.4} & 48.4 & 67.4 & 39.9 & 62.0 & 38.0 & 49.8 & \cellcolor{red!25}57.8 & \cellcolor{red!25}78.8\\
\midrule
jedit-4.1 & log4j-1.1  & 26.3 & 63.4 & 0.0 & 13.2 & 12.9 & \cellcolor{blue!25}\textbf{84.9} & 61.0 & 61.8 & \cellcolor{blue!25}\textbf{71.0} & 84.3 & 37.8 & 61.1 & \cellcolor{red!25}59.6 & 49.1 & 31.6 & 24.2 & 57.1 & \cellcolor{red!25}67.9\\
lucene-2.2 & log4j-1.1  & 64.1 & 74.1 & 60.4 & \cellcolor{blue!25}\textbf{92.0} & \cellcolor{blue!25}\textbf{70.8} & 82.3 & 52.0 & 60.9 & 63.0 & 79.7 & 45.2 & 53.8 & 46.5 & 41.7 & 53.2 & 62.4 & \cellcolor{red!25}54.5 & \cellcolor{red!25}63.1\\
\midrule
lucene-2.2 & xalan-2.5  & 63.6 & 57.4 & \cellcolor{blue!25}\textbf{68.5} & 60.9 & 61.7 & \cellcolor{blue!25}61.1 & 58.0 & 54.8 & 45.0 & 53.0 & 57.2 & 61.0 & 37.8 & 54.4 & 43.2 & 48.9 & \cellcolor{red!25}67.4 & \cellcolor{red!25}\textbf{66.6}\\
xerces-1.3 & xalan-2.5  & 38.4 & 50.7 & \cellcolor{blue!25}62.8 & \cellcolor{blue!25}59.0 & 21.8 & 56.0 & 59.0 & 53.9 & 57.0 & 53.5 & 26.8 & 46.9 & \cellcolor{red!25}\textbf{64.9} & 40.2 & 23.6 & 55.7 & 63.5 & \cellcolor{red!25}\textbf{61.1}\\
\midrule
xalan-2.5 & lucene-2.2  & 46.5 & 54.8 & \cellcolor{blue!25}\textbf{74.7} & \cellcolor{blue!25}\textbf{64.0} & 51.6 & 59.7 & 64.0 & 63.1 & 54.0 & 57.9 & 56.4 & \cellcolor{red!25}60.5 & \cellcolor{red!25}74.5 & 50.7 & 65.4 & 54.8 & 64.3 & 56.6\\
log4j-1.1 & lucene-2.2  & 49.3 & 62.3 & 37.8 & 57.9 & 55.0 & 60.8 & \cellcolor{blue!25}60.0 & 55.6 & 54.0 & \cellcolor{blue!25}\textbf{63.1} & 52.7 & 55.8 & \cellcolor{red!25}\textbf{76.4} & 56.1 & 62.9 & 50.0 & 70.2 & \cellcolor{red!25}\textbf{63.1}\\
\midrule
xalan-2.5 & xerces-1.3  & 35.4 & 55.9 & 0.0 & 37.8 & \cellcolor{blue!25}39.3 & \cellcolor{blue!25}64.7 & 23.0 & 39.5 & 31.0 & 49.6 & 32.4 & 57.5 & 15.7 & 43.4 & 34.4 & 62.4 & \cellcolor{red!25}\textbf{50.0} & \cellcolor{red!25}\textbf{74.3}\\
ivy-2.0 & xerces-1.3  & 12.5 & 64.2 & 0.0 & 33.8 & 20.0 & 52.7 & \cellcolor{blue!25}45.0 & \cellcolor{blue!25}66.7 & 37.0 & 60.3 & 36.6 & 59.6 & 29.4 & 51.3 & 32.1 & 53.2 & \cellcolor{red!25}\textbf{47.8} & \cellcolor{red!25}\textbf{71.7}\\
\midrule
xerces-1.3 & ivy-2.0  & 34.6 & 71.1 & \cellcolor{blue!25}\textbf{39.5} & \cellcolor{blue!25}\textbf{79.7} & 35.5 & 70.1 & 30.0 & 68.9 & 34.0 & 77.2 & 30.5 & 57.2 & 11.3 & 54.5 & 25.3 & 67.8 & \cellcolor{red!25}37.4 & \cellcolor{red!25}79.5\\
synapse-1.2 & ivy-2.0  & 33.3 & 74.3 & \cellcolor{blue!25}\textbf{51.1} & 78.7 & 34.7 & 74.5 & 24.0 & 62.5 & 38.0 & \cellcolor{blue!25}\textbf{82.1} & 29.6 & 62.0 & 22.0 & 18.2 & \cellcolor{red!25}40.7 & 71.9 & 39.0 & \cellcolor{red!25}78.9\\
\midrule
ivy-1.4 & synapse-1.1  & 9.4 & 66.1 & 20.9 & 35.6 & 3.4 & 63.2 & 45.0 & 61.4 & \cellcolor{blue!25}\textbf{51.0} & \cellcolor{blue!25}\textbf{70.0} & 9.7 & 51.9 & 15.7 & 54.1 & 9.4 & 37.4 & \cellcolor{red!25}42.7 & \cellcolor{red!25}57.8\\
poi-2.5 & synapse-1.1  & 28.3 & 48.1 & 34.9 & 54.2 & \cellcolor{blue!25}46.5 & \cellcolor{blue!25}62.9 & 43.0 & 62.7 & 5.0 & 44.4 & \cellcolor{red!25}\textbf{49.0} & 63.4 & 35.2 & 30.0 & 37.0 & 56.4 & 48.5 & \cellcolor{red!25}\textbf{68.6}\\
\midrule
ivy-2.0 & synapse-1.2  & 39.7 & 69.7 & 34.5 & 49.7 & 24.2 & 68.9 & 52.0 & 62.3 & \cellcolor{blue!25}57.0 & \cellcolor{blue!25}70.7 & 32.4 & 53.6 & 45.7 & 39.8 & 17.5 & 50.2 & \cellcolor{red!25}\textbf{62.0} & \cellcolor{red!25}\textbf{73.3}\\
poi-3.0 & synapse-1.2  & \cellcolor{blue!25}56.3 & \cellcolor{blue!25}69.9 & 55.8 & 66.0 & 53.8 & 56.2 & 56.0 & 67.6 & 43.0 & 62.8 & 49.5 & 62.3 & 29.4 & 34.0 & 49.8 & 55.2 & \cellcolor{red!25}\textbf{65.7} & \cellcolor{red!25}\textbf{75.1}\\
\midrule
synapse-1.2 & poi-3.0  & 57.7 & 74.1 & 51.7 & 59.3 & 27.2 & 70.0 & \cellcolor{blue!25}72.0 & 61.6 & 71.0 & \cellcolor{blue!25}75.6 & 48.5 & 59.5 & 73.9 & 56.5 & 66.2 & 56.7 & \cellcolor{red!25}\textbf{81.4} & \cellcolor{red!25}\textbf{82.2}\\
ant-1.6 & poi-3.0  & 47.0 & 68.5 & 47.2 & 53.2 & 37.8 & 70.2 & 38.0 & 33.9 & \cellcolor{blue!25}65.0 & \cellcolor{blue!25}79.7 & 43.5 & 66.0 & 33.3 & 56.4 & 44.7 & 41.6 & \cellcolor{red!25}\textbf{81.1} & \cellcolor{red!25}\textbf{84.3}\\
\midrule
\multicolumn{2}{c|}{Avg} &40.3 & 64.6 & 39.5 & 55.8 & 37.6 & 67.2 & 43.6 & 55.9 & \cellcolor{blue!25}48.9 & \cellcolor{blue!25}68.4 & 39.9 & 59.9 & 40.7 & 44.1 & 37.6 & 54.5 &  \cellcolor{red!25}\textbf{55.2} &  \cellcolor{red!25}\textbf{70.3}\\
\bottomrule
\end{tabular}
}
\end{table*}

\subsection{Answer to RQ3}
The cross-project defect prediction (CPDP) task mentioned in RQ3 mainly evaluates whether the contextual features learnt by the model can be applied to different projects.
To answer this question, we compared our proposed method, MFGNN, with several typical CPDP methods, and the results are shown in Table~\ref{table:promise_cpdp}. The best performance among all methods are marked in bold. Depending on the type of input data, we can further divide the performance into two types: the best performance among metric-based methods is marked in \colorbox{blue!25}{blue} and among source code model-based methods is marked in \colorbox{red!25}{red}.

% Generally speaking \he{or Theoretically?}, the results of CPDP should show consistency~\cite{consisten-cpdp}, i.e., the best performing source project on a target project is the same for all methods.\he{cannot understand the above sentence}
% For two of the target projects (synapse-1.1 and poi-3.0), the property holds for the methods with a source code model (e.g., DTLDP, MFGNN, etc.). However, the property never holds for the traditional methods with software metrics. This is because the software metrics are too rigid to reflect the versatile source code features.

Among all methods, MFGNN achieved the highest overall F1 and AUC. Compare to the best metric-based methods, MFGNN outperformed 6.3\% and 1.9\% in F1 and AUC, respectively. Compare to other source code model-based methods, MFGNN achieved the highest F1 and AUC in most of the tasks.
Interestingly, the BugContext does not perform as well as its result on the WPDP task (see Section~\ref{sec:answer-rq2}). Compared with BugContext, the F1 and AUC of MFGNN were improved by up to 27.6\% and 15.8\%, respectively. 
We think the reason of improvement lies behind their difference of using context-dependent information, which could be divided into two-fold. On one side, BugContext uses dependency features to assist AST features, while MFGNN does the opposite. Learning program context-dependent features is critical for CPDP task, thus such a design difference can lead to a discrepancy in performance. 
On the other side, BugContext extracts features from CFG and DFG separately, while MFGNN combines them into the \zzhb{ECFG} and extracts features uniformly by AGN4D.

In conclusion, the contextual features obtained by MFGNN are more generalized and are able to result in better performance on the CPDP task.

\subsection{Answer to RQ4}
\zzh{Table~\ref{tab:ccd} illustrates the results related to RQ4, and the best results are highlighted in bold. Compared with other methods, MFGNN achieved the highest recall and F1, as well as a relative high precision.
Interestingly, we could observe that FCDetect plays well, which apply call graph as the source code model. However, we argue that MFGNN can capture the program context-dependency features more effectively. The main difference between them is the graph learning mechanisms they adopted. Compared to the Graph2Vec adopted by FCDetect, MFGNN uses AGN4D based on the attention mechanism, and thus could \bo{adjust the weights of different types of dependency information}.
Therefore, with the help of more context-dependency features, MFGNN could identify program variants more effectively, leading to higher recall and F1 scores.
% Except for the context-dependency features, MFGNN can also capture \zzhb{function}-dependent features more effectively.
% That is because both FCDetect and MFGNN have high recall compared to the remaining methods. For the current task, i.e., functional code clone detection, the higher recall means that the methods are sensitive to functional changes, which is highly correlated with \zzhb{function-dependent features}. This indicates that the MFGNN can effectively capture \zzhb{function}-dependent features.
\begin{table*}[!h]
    \centering
    \caption{The results of ccd task on OJClone.} \label{tab:ccd}
    \begin{tabular}{cccccccc}
    \hline
    Methods & RAE+ & Deckard & CDLH & ASTNN & DeepSim & FCDetect & MFGNN\\
    \hline
    P       & 52.5 & \textbf{99} & 47   & 98.9  & 70 &  97 & 96.7 \\ 
    R       & 68.3 & 5 & 73   & 92.7  & 83 & 95 & \textbf{96.3} \\
    F1      & 59.4 & 10 & 57   & 95.5  & 76& 96 & \textbf{96.5} \\
    \hline
\end{tabular}
\end{table*}

\begin{figure}[!h]
    \includegraphics[scale=0.4]{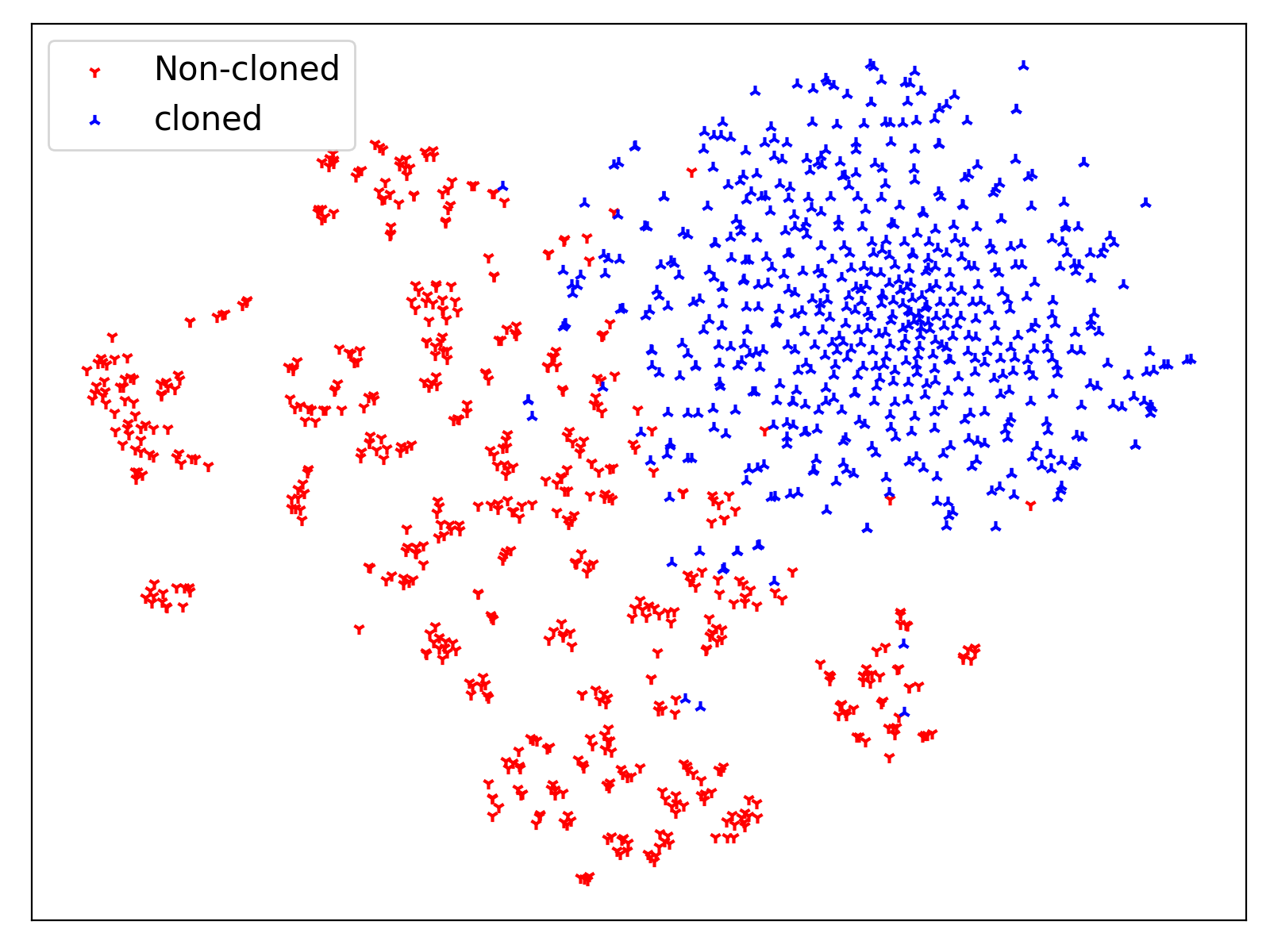}
    \caption{t-SNE mapping of the absolute distances of the test set's pairs' features.}\label{fig:ccd}
\end{figure}

\zzhb{Figure~\ref{fig:ccd} shows the absolute distances of features derived from MFGNN for the data in the test set. 
We can observe that there is a clear demarcation line between the red and blue dots.
This illustrates the features obtained by MFGNN can effectively distinguish source codes under the functional code-clone task.}
In conclusion, MFGNN can improve the performance of distinguishing between non-cloned and cloned source code pairs.
}

\subsection{Answer to RQ5}
To answer this question, we have adjusted the default settings in our original methodology and compared their performance on the program classification task. The results are shown in Table~\ref{tab:ablation}, in which \zzh{the default settings are highlighted in bold.} We can obtain the following insights:

\textbf{\textit{Sensitivity to the control flow and dataflow differs from challenges.}} Using only DFG as the source code model~(i.e., A+D+S) works better on some challenges, e.g., \texttt{SUB} and \texttt{SUM}. 
\zzhb{This is because there are much more operators than branches within these source code (see Table \ref{tab:statistic}). In other words, these challenges have simple control flows, but complex computational logic, which is related to data flow heavily. Thus, compare to control flow edges, data flow edges play a more critical role on the test results.
However, in general, CFG only~\zzh{(i.e., A+C+S)} could perform better than adopting only DFG.}

\textbf{\textit{Introducing different types of edges in CFG may lead to poorer performance.}} 
Introducing different types of edges plays a positive role on some challenges, including \texttt{SUB}, \texttt{721C}, \texttt{731C}, \texttt{742C}, and \texttt{822C}. However, on other challenges, MFGNN performs better when the source code model is untyped (e.g., A+C+S). \zzhb{This is because these challenges require fewer branches than the others (see Table~\ref{tab:statistic}). 
The imbalanced distribution of types lead to ineffective optimization of the model on different types.
Therefore, the uneven distribution of the number of different edge types prevents MFGNN from effectively fusing the features of different types of flows.}

\begin{table*}[!h]
\centering
\setlength\tabcolsep{3pt}
%\small
\caption{Results of ablation studies}\label{tab:ablation}
\label{table:f1_ablation}
\resizebox{\textwidth}{!}{
\begin{tabular}{ccccccccccccccccccc|cc}
    \toprule
    \multirow{2}{*}{Different Settings} & \multicolumn{2}{c}{SUB} & \multicolumn{2}{c}{MNMX} & \multicolumn{2}{c}{FLOW} & \multicolumn{2}{c}{SUM} & \multicolumn{2}{c}{1062C} & \multicolumn{2}{c}{721C} & \multicolumn{2}{c}{731C} & \multicolumn{2}{c}{742C} & \multicolumn{2}{c}{822C} & \multicolumn{2}{|c}{Avg}\\
    & Acc & F1 & Acc & F1 & Acc & F1 & Acc & F1 & Acc & F1 & Acc & F1 & Acc & F1 & Acc & F1 & Acc & F1 & Acc & F1\\
    \midrule
    A+C+S&70.6 & 70.6 & 80.9 & 79.5 & 82.6 & 72.2 & 71.4 & 72.2 & 66.8 & 57.0 & 59.2 & 53.5 & 66.4 & 56.9 & 74.6 & 48.6 & 60.2 & 50.8 & 70.3 & 62.4\\
    A+D+S&70.8 & 71.4 & 80.9 & 76.7 & 78.7 & 69.3 & 71.5 & 71.8 & 66.2 & 43.5 & 57.5 & 53.7 & 63.8 & 56.3 & 72.2 & 45.0 & 56.3 & 49.6 & 68.7 & 59.7\\
    A+C+M&70.6 & 72.2 & 80.7 & 77.6 & 82.1 & 70.7 & 70.8 & 70.8 & 64.9 & 48.8 & 59.7 & 54.2 & 66.9 & 56.6 & 75.5 & 52.9 & 59.4 & 52.0 & 70.1 & 61.8\\
    B+C+D+M&69.7&68.3&75.4&68.8&72.8&63.5&64.6&64.0&57.1&38.8&49.0&44.2&61.6&51.7&68.0&40.3&55.7&43.9&63.8&53.7\\
    \textbf{A+C+D+M}&74.5 & 74.7 & 83.1 & 81.4 & 81.8 & 71.0 & 72.9 & 73.5 & 68.0 & 53.2 & 59.5 & 54.5 & 70.0 & 61.0 & 73.8 & 51.6 & 59.9 & 50.3 & \textbf{71.5} & \textbf{63.5}\\
    \midrule
    concatenation & 66.9 & 63.8 & 79.5 & 74.6 & 80.6 & 69.4 & 70.6 & 70.1 & 65.9 & 51.0 & 55.8 & 50.4 & 66.8 & 54.0 & 69.8 & 43.4 & 60.3 & 47.0 & 68.5 & 58.2\\
    \textbf{summation}&74.5 & 74.7 & 83.1 & 81.4 & 81.8 & 71.0 & 72.9 & 73.5 & 68.0 & 53.2 & 59.5 & 54.5 & 70.0 & 61.0 & 73.8 & 51.6 & 59.9 & 50.3 & \textbf{71.5} & \textbf{63.5}\\
    \midrule
    GCN&72.3 & 70.4 & 80.4 & 77.2 & 81.3 & 71.8 & 70.7 & 70.8 & 64.6 & 43.5 & 56.6 & 52.2 & 64.3 & 53.2 & 69.7 & 39.9 & 59.8 & 48.8 & 68.9 & 58.6\\
    GGNN&74.7 & 73.9 & 83.2 & 81.7 & 82.8 & 72.1 & 71.5 & 71.2 & 62.7 & 41.6 & 53.7 & 48.1 & 63.6 & 50.6 & 69.5 & 42.7 & 56.7 & 38.0 & 68.7 & 57.8\\
    \textbf{AGN4D}&74.5 & 74.7 & 83.1 & 81.4 & 81.8 & 71.0 & 72.9 & 73.5 & 68.0 & 53.2 & 59.5 & 54.5 & 70.0 & 61.0 & 73.8 & 51.6 & 59.9 & 50.3 & \textbf{71.5} & \textbf{63.5}\\
    \bottomrule
\end{tabular}
}
\end{table*}

\zzh{\textbf{\textit{AST is the a better choice for node representation in our experiment settings.}}
The results show that using AST as a node representation improved the model's performance significantly. Even when the other settings in the approach were removed (e.g., A+C+S which removed data flow edges and edge types, or A+D+S which removed control flow edges), the approach still performed better than B+C+D+M, which represents node by Bag-of-Words~(BoW) model instead of AST. Compared to the model in BoW, i.e., B+C+D+M, our source code model (A+C+D+M) resulted in 7.7\% and 9.8\% improvement on accuracy and F1, respectively. Because the node representation is the only independent variable here, we can conclude that AST is a better node representation option for our task. Compared to AST, BoW lacks both the lexical order and syntactic structures, which are essential for a proper representation of basic blocks.
}

\textbf{\textit{AGN4D is the best choice among the three GNNs}}. To examine the effectiveness of AGN4D, we altered it into two other common GNNs, i.e., GCN and GGNN, respectively, into our approach for a comparison study. Table~\ref{tab:ablation} shows that AGN4D outperformed the other two GNNs, with an average of 2.7\% and 5.3\% higher accuracy and F1, respectively.

\zzhb{\textbf{\textit{Summation is a better choice than concatenation in contextual feature embedding stage.}} From the results, the use of summation as a graph feature synthesis method (i.e., the last formula of \ref{eqn:2}) delivered better performance. This is because concatenation doubles AGN4D's hidden dimension layer by layer, increasing the number of model parameters and resulting in model overfitting issue.}

\subsection{Threats to Validity}
\zzh{
In conducting our experiments, the following factors existed that might affect the validity of the our study.

\textbf{Implementation of baselines.} The internal threat to validity is concerned with our implementation. We reproduced TBCNN, ASTNN, CtxG, DBN, TCA+, TreeLSTM, BugContext. Although we have implemented these baseline methods as described in the original studies, we cannot guarantee that these implementations exactly match the original ones. 

\textbf{Applying baselins on our dataset.} In carrying out the task, we found that many of the baseline methods were designed specifically for a particular task, for example code2vec's goal was to perform function name generation and CtxG's goal was to perform var-misuse detection. Although we compared these methods as baselines, we cannot guarantee that these we can meet the conditions for these representations of the model to work well.

\textbf{Missing projects in PROMISE dataset.} Our RQ2 and RQ3 experiments are based on the PROMISE dataset, a very early dataset in which some versions of projects recorded are not available on the web. We were only able to conduct experiments using projects that could be found and could not directly use the original experimental data from the DBN~\cite{url1} and TreeLSTM~\cite{url12} studies.  

\textbf{CFG differences in different languages.} For C/C++, we use Clang to get the CFG, which converts the program to LLVM IR, a kind of three-address code, and then builds the CFG on top of that. For Java, we use Soot to get the CFG. Soot will first convert the program into Jimple, a kind of SSA, and then build the CFG on top of that. Because of the difference in the intermediate languages used, the final CFG may not be exactly the same for the same statements in both languages.

\zzhb{\textbf{Conduct experiments on more tasks and more practical datasets.} To evaluate the feasibility and effectiveness of MFGNN, we have conducted several tasks (e.g., program classification and defect prediction) on the datasets consisting of source codes from OJ and open source projects. Though the variety of evaluated tasks and the sources of datasets were limited, we argue that MFGNN is robust enough even on large-scale real-world industrial code to perform other types of tasks, which, however, requires follow-up studies in the future.}
}

\section{Related Works}
\subsection{Source Code Representation in Deep Learning}
\zzhb{
While performing program analysis with deep learning, the representation model of source code is a fundamental problem, which could be roughly divided into: AST-based and CFG-based.
Specifically, as for the AST-based source code model, some studies adopted the AST that is generated from the program directly~\cite{url9,url10,url12} or with some modifications (e.g., inserting additional edges between nodes~\cite{url11}). Moreover, some works~\cite{url13, c2v, c2s} just extracted part of the generated AST to conduct the following analysis. For examples, Alon et al.~\cite{c2v, c2s} chose the collection of AST's token-to-token path as the source code model, and learned the features by attention-based models.
Unlike these models, we chose to split the AST into subtrees based on basic blocks. Though it would slightly broke the integrity of the AST, the explicit contextual dependencies in the CFG could reassemble parts of the AST, making dependencies more salient and easier to learn.

As for the CFG-based source code model, there are two factors that significantly affect the following program analysis with deep learning. One is the way of representing of basic blocks; the other is the role of the graph. To be specific, several works have tried different way to basic blocks in deep learning, e.g., assembly instruction~\cite{url14}, Bag-of-Words model~\cite{FCD, ginn} and line number~\cite{Li2019ImprovingBD}. As for the graph, it can be utilized as a leading role~\cite{url14, FCD, ginn} or an auxiliary role~\cite{Li2019ImprovingBD} during the analysis. For example, Wang et al.~\cite{ginn} used graph as a leading role and represents basic blocks with Bag-of-Words model composed of AST's grammatical nodes. Our model similarly adopted graph as a leading role, but represented basic blocks with the corresponding subtree of AST. We retained the structure of AST, which helped us better represent the context-independent grammatical differences than other models.
}

\subsection{Program Classification}
\zzhb{
Program classification, i.e., distinguishing and classifying programs by some features from various aspects, is one of the basic software engineering tasks. 
For example, as one of the applications, functional code clone detection~\cite{url13, FCD, tbcnnccd} is to determine whether two code snippets implement the same functionality. It is achieved by classifying the functional features of the given program.
Except from functional features~\cite{url10, url13}, language features~\cite{url21}, defect features~\cite{url12, url1, url14} and structure features~\cite{url20} are also widely adopted by program classification tasks.
In this paper, we decided to apply defect features on classifying program test results. Though Phan et al.~\cite{url14} have done this task before, the size of dataset and code complexity were relatively limited compared to ours, which were collected and constructed by crawlers and huge manual efforts.
}

\subsection{Software Defect Prediction}
\zzhb{
Software defect prediction is a challenging task that has been researched extensively.
Prior to the rise of deep learning, researchers have adopted machine learning to achieve such a goal~\cite{url22,url23,url24,url25,url38,url41, url40, url36,url39,url37}. However, these techniques require feature engineering that is normally time- and resource-consuming. For example, Xing et al.~\cite{url39} proposed a SVM-based defect predictin methods, which depends on both software change metrics and software complexity metrics. 
Deep learning techniques eliminated the process of feature engineering, and researchers began focusing on improving prediction performance using suitable source code models~\cite{url23, url1, DTLDP, url12}.
Existing works have pointed out that the source code model needs contextual dependencies~\cite{Li2017SoftwareDP} and should be able to distinguish subtle changes~\cite{url1}. Both of them were taken into account in our method. Specifically, the contextual dependencies comes from the ECFG; and the subtle changes, i.e., subtle grammatical differences, are represented by the structural differences of the AST.
To the best of our knowledge, no other existing source code models have achieved both of these goals. 
}

\section{Conclusion}
\zzhb{In this paper, we have proposed a new source code model based on ECFG and an attention-based model, namely MFGNN. 
Our source code model restricts the order in which MFGNN extracts features, and makes it more efficient and effective for MFGNN to obtain program features.
Moreover, we have evaluated MFGNN on three practical tasks: program classification, software defect prediction and code clone detection. The results showed that MFGNN significantly outperformed baseline methods. For example, compared with the well-known source code model code2seq~\cite{c2s}, the scale of parameters decreased more than 30-fold while the overall accuracy was increased by 26.0\%.
Our research illustrated that the performance heavily depended on the construction of source code model. Additionally, we highlights a few research directions for future work, e.g., applying our method on more general real-life projects and improving the graph and MFGNN for better performance.}

\section*{Acknowledgement}
We thank anonymous reviewers for their thoughtful comments. Thanks to Ningyu He for proof- reading the manuscript. This research is supported by the National Key R\&D Program of China under Grant No. 2020AAA0109400, the National Natural Science Foundation of China under Grant Nos. 62072007, 61832009, 61620106007, 61502011, the Australian Research Council Discovery Project (Grant No. DP210102447), and "the Fundamental Research Funds for the Central Universities"(BLX202003).

%% The Appendices part is started with the command \appendix;
%% appendix sections are then done as normal sections
%% \appendix

%% \section{}
%% \label{}

%% References
%%
%% Following citation commands can be used in the body text:
%% Usage of \cite is as follows:
%%   \cite{key}         ==>>  [#]
%%   \cite[chap. 2]{key} ==>> [#, chap. 2]
%%

%% References with BibTeX database:

\bibliographystyle{elsarticle-num}
\bibliography{reference}

%% Authors are advised to use a BibTeX database file for their reference list.
%% The provided style file elsarticle-num.bst formats references in the required Procedia style

%% For references without a BibTeX database:

% \begin{thebibliography}{00}

%% \bibitem must have the following form:
%%   \bibitem{key}...
%%

% \bibitem{}

% \end{thebibliography}

\end{document}